\documentclass{article}
\usepackage{amssymb}
\usepackage{amsmath}
\usepackage{amsfonts}
\usepackage{graphicx}

\let\UnmodifSec=\section
\renewcommand{\section}{\setcounter{equation}{0}\UnmodifSec}

\newtheorem{lemma}{Lemma}[section]

\newtheorem{theorem}{Theorem}[section]
\newtheorem{remark}{Remark}[section]

\renewcommand{\cosh}{\mbox{ch\,}}\renewcommand{\sinh}{\mbox{sh\,}}

\def\bC{{\bf C}}
\def\bR{{\bf R}}
\def\bN{{\bf N}}
\def\bZ{{\bf Z}}

\def\Im{\mathop{\rm Im}\nolimits}
\def\Re{\mathop{\rm Re}\nolimits}

\def\ch{\mathop{\rm ch}\nolimits}
\def\sh{\mathop{\rm sh}\nolimits}

\def\AA{{\cal A}}

\def\CC{{\cal C}}

\def\FF{{\cal F}}

\def\LL{{\cal L}}

\def\TT{{\cal T}}

\def\WW{{\cal W}}

\def\wh{\widehat}
\def\wt{\widetilde}
\def\ovl{\overline}

\def \vhi{\varphi}
\def \veps{\varepsilon}

\def\half{{\scriptstyle{1 \over 2}}}
\def\Beta{{\rm B}}
\def\interior#1{\setbox1=\hbox{$#1$}\rlap{$#1$}\kern0.4\wd1\raise1.1\ht1%
\hbox{$\scriptstyle \circ$}}
\def\bydef{\mathrel{\buildrel \hbox{\scriptsize \rm def} \over =}}
\def\boxit#1#2{\setbox1=\hbox{\kern#1{#2}\kern#1}%
\dimen1=\ht1 \advance \dimen1 by #1 \dimen2=\dp1 \advance \dimen2 by #1
\setbox1=\hbox{\vrule height\dimen1 depth\dimen2\box1\vrule}%
\setbox1=\vbox{\hrule\box1\hrule}%
\advance \dimen1 by .4pt \ht1=\dimen1 \advance \dimen2 by .4pt \dp1=\dimen2
\box1\relax}
\def\endprf{\raise .5ex\hbox{\boxit{2pt}{\ }}}
\def\thvbar{\mathrel{\vrule height 2ex depth 0pt width 0.1 em}}

\def\ifundefined#1{\expandafter\ifx\csname#1\endcsname\relax}

\def\beq{\begin{equation}}
\def\endq{\end{equation}}
\def\beqa{\begin{eqnarray}}
\def\endqa{\end{eqnarray}}

\def\bal{\begin{align}}
\def\endali{\end{align}}

\def\coupl{{\gamma}}

\def\bin#1#2{\left ({{#1}\atop{#2}} \right )}

\def\P{{\bf P}}
\def\D{{\bf D}}
\def\Q{{\bf Q}}

\begin{document}

\title{Anti de Sitter quantum field theory and a new class of 
hypergeometric identities}
\author{Jacques Bros$^1$, Henri Epstein$^2$, Michel Gaudin$^1$,
Ugo Moschella$^{3,4}$,
Vincent Pasquier$^1$ \\
\\$^1$Institut de Physique Th\'eorique,
CEA - Saclay, France\\
$^2$Institut des Hautes \'Etudes
Scientifiques, 91440 Bures-sur-Yvette \\
 $^3$Universit\`a dell'Insubria, 22100 Como,
Italia,
\\$^4$INFN, Sez. di Milano, Italia}
\maketitle
\begin{abstract}
We use Anti-de Sitter quantum field theory to prove a new class of identities 
between hypergeometric functions related to the K\"all\'en-Lehmann 
representation of products of two Anti-de Sitter two-point functions. 
A rich mathematical structure emerges. We apply our results  to study 
the decay of unstable Anti-de Sitter particles. The total amplitude is 
in this case finite and Anti-de Sitter invariant.
\end{abstract}
\section{Introduction}
\label{intro}
The interest in the
Anti-de Sitter geometry and the corresponding classical and 
quantum field theories has gradually increased
in recent years and gained an important place in theoretical physics.
Today, studies in Anti-de Sitter field theory or researches
using Anti-de Sitter techniques to compute amplitudes in other kind of 
(realistic) quantum field theories such as quantum chromodynamics
play a central role in high energy physics.

Anti-de Sitter provides indeed access to nontrivial Minkowski 
quantum field theories in two ways.
Through the Maldacena duality \cite{maldacena}, Anti-de Sitter models 
correspond to conformal quantum field theories on the boundary.
In this approach quantum theories on the Minkowski spacetime come  
from (and actually are believed to be equivalent to)
models in higher dimensional Anti-de Sitter universes.

On the other hand, the Anti-de Sitter manifold may be also viewed as an infrared
(covariant) regularization of the Minkowski spacetime \cite{callan};
Poincar\'e invariant models can be constructed by taking the flat limit 
of Anti-de Sitter ones.
In this way one can gain information on Minkowskian quantum field theories
from Anti-de Sitter models having the same spacetime dimensionality.

In both cases, the correspondences between Anti-de Sitter and Minkowski
theories may be used to uncover new pieces of mathematics.
The idea is that to a known relation existing on the Minkowski spacetime 
there should correspond a possibly unknown relation
on the Anti-de Sitter universe and vicecersa.

In this paper we use this idea to guess and prove a new class of 
linearization identities among hypergeometric functions.
This was suggested by  a series of related papers \cite{bros3,bros2,bros1} 
where we have considered particle decays in the de Sitter universe.
The effort necessary to compute the K\"all\'en-Lehmann weights needed to 
evaluate the lifetime of de Sitter particles
unveiled there a rich mathematical structure;  new integral formulae for 
products of three Legendre functions followed.

Trying to solve the same problem in the Anti-de Sitter case provides
a new class of nontrivial identities between hypergeometric functions.
The mathematics behind these new identities is however quite different.

In the end of the paper, as an application of our results, we briefly discuss 
the problem of particle decay in the Anti-de Sitter universe and its 
flat limit. This example will make clear the value of the Anti-de Sitter 
universe as an infrared regulator of calculations which are divergent in the 
flat  case.
In particular we compute the total probability of decay  of a given 
Anti-de Sitter
one-particle state into all possible two-particle states at first order in 
perturbation theory.
This quantity is divergent both in the Minkowski and the de Sitter universes 
while it is perfectly finite and can be explicitly computed in the 
Anti-de Sitter case. This quantity, once it is divided by the radius of 
the Anti-de Sitter universe, has a flat limit proportional to
the inverse lifetime of a corresponding Minkowski unstable particle as 
it is usually computed by means of the Fermi golden rule.
However we have not yet fully solved the problem of finding a unique 
Anti-de Sitter normalization
to get the right dependence of such lifetime on the speed of the 
Minkowski particle.
The point is that the lifetime is obtained in the Minkowski case as the 
ratio of two divergent quantities while the Anti-de Sitter amplitude is 
already finite and it is not completely clear what the 
"amplitude per unit time" should be in the Anti-de Sitter spacetime. 
It is an interpretation problem that  we leave for further investigation.

Section \ref{prelim} recalls some well-known facts and fixes some notations.
Sections \ref{exprob}, \ref{alg}, and \ref{dim3} give a precise statement
and preliminary discussions of the main mathematical problem
to be solved in this paper, and Section \ref{ana} gives its solution.
Section \ref{expthms} applies this result to expansion theorems
for second kind Gegenbauer functions and the K\"all\'en-Lehmann
expansion of the product of two free-field two-point functions
in AdS (or its covering). Section \ref{appli} gives the applications
to quantum field theory in AdS mentioned above.

\section{Preliminaries}
\label{prelim}
The $d$-dimensional real and complex Anti-de Sitter (AdS) space-times
with radius $R>0$ are respectively defined as
\beq
X_d = \{x \in \bR^{d+1}\ :\ x\cdot x = R^2\}\ \ \ \ {\rm and}
\ \ \ \ X_d^{(c)} = \{x \in \bC^{d+1}\ :\ x\cdot x = R^2\},
\label{s.1}\endq
where the scalar product $x\cdot y$ is defined as
\beq
x\cdot y = x^0 y^0 + x^d y^d - x^1 x^1 - \dots\ -x^{d-1} x^{d-1}\ \ =\ \
x^0 y^0 + x^d y^d - \vec{x}\cdot \vec{y}\ .
\label{s.2}\endq
The vector $e_\mu \in \bR^{d+1}$ has coordinates
$e_\mu^\nu = \delta_{\mu \nu}$.
$G_0$ (resp. $G_0^{(c)}$) is the connected component of the unit
in the group of real (resp. complex) linear transformations of
$\bR^{d+1}$ (resp. $\bC^{d+1}$) which preserve the scalar product
(\ref{s.2}). The future and past tuboids $\TT_{1\pm}$ are given by
\beq
\TT_{1+} = (\TT_{1-})^* = \{x+iy \in X_d^{(c)}\ :\ y\cdot y > 0,\ \ \
y^0 x^d - y^d x^0 > 0\}.
\label{s.3}\endq
These tuboids are invariant under $G_0$. Their properties
are studied in detail in \cite{brosads}. The universal covering spaces of
$X_d$, $G_0$, $\TT_{1\pm}$ are respectively denoted
$\wt X_d$, $\wt G_0$, $\wt \TT_{1\pm}$.  We will assume $d \ge 2$.
In this paper, we will take $R=1$ except when it is explicitly
stated otherwise.

We denote:
\beq
\bC_+ = - \bC_- = \{z\in \bC\ :\ \Im z >0\},\ \ \ \
\Delta_1 = \bC \setminus [-1,\ 1],\ \ \ \
\wh\Delta_1 = \Delta_1 \cup \{\infty\},
\label{s.4}\endq
(i.e. $\wh\Delta_1$ is the complement of the segment $[-1,\ 1]$
in the Riemann sphere), and $\wt \Delta_1$ the universal
cover of $\Delta_1$. The image of $\TT_{1-}\times\TT_{1+}$
given by the map $(z_1,\ z_2) \mapsto z_1\cdot z_2$ is
$\Delta_1$.

A function $f$ holomorphic in
$\bC_+\cup \bC_-$ is said to have tempered behavior if there are
positive integers $M$, $P$ such that
\beq
\thvbar f \thvbar_{M,P} \bydef
\sup_z (1+|z|)^{-M}\,(1+|\Im z|^{-1})^{-P} |f(z)|
<\infty\ .
\label{s.4.1}\endq
Such a function has
boundary values $f_+$ and $f_-$ on the real axis in the sense
of tempered distributions from $\bC_+$ and $\bC_-$ respectively,
and we denote ${\rm disc}\,f = f_+-f_-$. If $T$ is a tempered
distribution on $\bR$ with sufficient decrease at infinity
(in particular if it has compact support) then
\beq
f(z) = {1\over 2\pi i} \int_{\bR} {T(t)\,dt\over t-z}
\label{s.5}\endq
is holomorphic with tempered behavior in $\bC_+\cup\bC_-$
and ${\rm disc}\,f = T$. A sequence $f_n$ of functions
holomorphic in $\bC_+\cup\bC_-$ tends to 0 in the sense of
functions with tempered behavior if there are
positive integers $M$, $P$ such that
$\thvbar f_n \thvbar_{M,P}\, \rightarrow 0$.
In this case $f_{n\pm} \rightarrow 0$ in the sense of tempered distributions.
A function $f$ holomorphic in $\TT_{1\pm}$ is said to have
tempered behavior if there are
positive integers $M$, $P$ such that
\beq
\thvbar f \thvbar_{M,P} \bydef
\sup_{z= x+iy \in \TT_{1\pm}} (1+|z|)^{-M}\,(1+|y\cdot y|^{-1})^{-P} |f(z)|
<\infty\ .
\label{s.5.1}\endq

If $\phi$ is a neutral scalar local quantum field on $X_d$
satisfying standard assumptions (see \cite{brosads,sw}), there is a function
$W$ holomorphic in $\TT_{1-}\times \TT_{1+}$, and a function
$w$ holomorphic with tempered behavior in $\Delta_1$, such that,
in the sense of tempered distributions,
the two-point vacuum expectation value of $\phi$ satisfies
\begin{align}
&\WW(x_1,\ x_2) \bydef (\Omega,\ \phi(x_1)\phi(x_2)\,\Omega) =
\lim_{\begin{array}{c}
\scriptstyle z_1 \in \TT_{1-},\ \ z_2 \in \TT_{1+}\\
\scriptstyle z_1\rightarrow x_1,\ \ z_2\rightarrow x_2 \end{array}}
W(z_1,\ z_2)\ ,
\label{s.6}\\
&W(z_1,\ z_2) = w(z_1\cdot z_2)\ .
\label{s.7}\end{align}
Conversely, if $w$ is a function holomorphic with tempered behavior in
$\Delta_1$, there exists a generalized free field $\phi$ such
that (\ref{s.6}, \ref{s.7}) hold (it will satisfy the positivity condition
if and only if $(z_1,\ z_2) \mapsto w(z_1\cdot z_2)$ is of
positive type). In the case of $\wt X_d$, $w$ is replaced by
a function holomorphic on $\wt \Delta_1$; we will mostly
consider its restriction to the cut-plane $\bC\setminus (-\infty,\ 1]$.

In the special case of the standard scalar neutral Klein-Gordon field
with mass $m$ on $X_d$, each of the functions $\WW$, $W$, and $w$
is labelled by a parameter $\nu$
of the form $\nu = n+(d-1)/2$, where $n$ is an integer $n > (1-d)$,
related to the mass by
\beq
m^2 = n(n+d-1)\ .
\label{s.8}\endq
The function $w_{n+{d-1\over 2}}$ is given by
\begin{align}
w_{n+{d-1\over 2}}(z) &=
{e^{-i\pi{d-2\over 2}}\over (2\pi)^{d\over 2}} (z^2-1)^{-\frac{d-2}4}
Q_{n+{d-2\over 2}}^{d-2\over 2}(z)
\label{s.9}\\
&=
{\Gamma \left ({d-1\over 2} \right )\over
2\pi^{d+1\over 2}} D_n^{d-1\over 2}(z)\ .
\label{s.10}\end{align}
Here and in the sequel $z^\alpha = \exp(\alpha \log z)$ is defined
as holomorphic in $\bC \setminus \bR_-$ and $(z^2-1)^\alpha$ as
$z^{2\alpha}(1-z^{-2})^\alpha$. The function $z \mapsto (1-z^{-2})^\alpha$
is holomorphic in $\wh \Delta_1$.
The function $Q_\alpha^\beta$ is the Legendre function of the second kind
(see \cite[pp.~122 ff]{HTF1} ) which is defined for complex values
of $\alpha$ and $\beta$, and the function $D_n^\lambda$, a Gegenbauer
function of the second kind, is also defined for complex values
of $n$ and $\lambda$. The following formulae will play an important
role in this paper:
\begin{eqnarray}
D_n^\lambda(z) &=& {\pi\Gamma(n+2\lambda)\over
\Gamma(\lambda)\Gamma(n+\lambda+1)} (2z)^{-n -2\lambda}
F \left ( {n+2\lambda\over 2},\ {n+2\lambda+1\over 2}\ ;\
 n+\lambda+1\ ;\ {1\over z^2} \right )
\label{s.11}\\
&=& {\pi\Gamma(n+2\lambda)\over
\Gamma(\lambda)\Gamma(n+\lambda+1)} (\zeta)^{-n -2\lambda}
F \left ( n+2\lambda,\ \lambda\ ;\
 n+\lambda+1\ ;\ {1\over \zeta^2} \right )\ ,
\label{s.12}
\end{eqnarray}
where the variables $z$ and $\zeta$ are related as follows
\begin{eqnarray}
& & \zeta = z+(z^2-1)^\half,\ \ \ \zeta^{-1} = z-(z^2-1)^\half,\ \ \ \
z = {\zeta+\zeta^{-1}\over 2}\ .
\label{s.13}
\end{eqnarray}
The above formulae do not require any of the parameters to be an integer,
but we will always assume $\Re(n+2\lambda) >0$ when using them.
The equality of (\ref{s.11}) and (\ref{s.12}) is explained
in Appendix \ref{hypid}.
The functions $D_n^{\lambda}$ are further discussed in Appendix \ref{jac}.
Formulae (\ref{s.9}, \ref{s.10}) extend, mutatis mutandis,
to the covering $\wt X_d$ of the Anti-de Sitter spacetime, but then $n$ is not any longer required to be an integer.

We denote $E(L)$, with $L > 1$, the ellipse with foci
$\pm 1$ given by
\beq
E(L) = \{\half(\zeta+\zeta^{-1})\ :\ \zeta \in \bC,\ \ |\zeta| = L \}.
\label{s.14}\endq
The outside $E_+(L)$ and inside $E_-(L)$ of $E(L)$ are defined by
\begin{align}
&E_+(L) = \{\half(\zeta+\zeta^{-1})\ :\ \zeta \in \bC,\ \ |\zeta| > L \},\cr
&E_-(L) = \{\half(\zeta+\zeta^{-1})\ :\ \zeta \in \bC,\ \
1< |\zeta| < L \} \cup [-1,\ 1]\ .
\label{s.15}\end{align}
We also define $E_+(1) = \Delta_1$. Note that if $z$ and $\zeta$
are related by (\ref{s.13}) then $z^{-2}$ is expressible as a
series in powers of $\zeta^{-2}$ which converges for
$|\zeta| >1$ and vice-versa.

We will frequently use the classical notation
$(t)_k = t(t+1)\ldots(t+k-1) = \Gamma(t+k)/\Gamma(t)$ if $k$ is
an integer $\ge 1$, $(t)_k = 1$ if $k \le 0$.

We will also use the notation
\beq
\alpha_\lambda(s)
 =\ \ \frac{\Gamma(s+\lambda)}{\Gamma(\lambda)\Gamma(s+1)} \ \
= \frac{1}{s\,\Beta (\lambda,s)}
\label{s.16}\endq

\section{The expansion problem}
\label{exprob}

If $\WW_m(x_1,\ x_2)$ denotes the two-point vacuum expectation value of a
free neutral scalar Klein-Gordon quantum field with mass $m$
on Minkowski space-time,
and if $F(x_1,\ x_2)$ is any function with the same general linear
properties as the two-point function of a local field, there
exists a tempered weight $\rho$ with support in the positive real axis
such that
\beq
F(x_1,\ x_2) = \int_{\bR_+} \rho(m^2)\,\WW_m(x_1,\ x_2)\,dm^2\ .
\label{i.1}\endq
$\rho$ is called the K\"all\'en-Lehmann weight associated to $F$,
and it is a positive measure if and only if $F$ is of positive
type (see e.g. \cite[p.~336]{BLOT}).
In particular for any two given masses $m_1$ and $m_2$
\beq
\WW_{m_1}(x_1,\ x_2)\,\WW_{m_2}(x_1,\ x_2) =
\int_{(m_1+m_2)^2}^\infty \rho_{\rm Min}(a^2;\ m_1,\ m_2)\,
\WW_{a}(x_1,\ x_2)\,da^2\ ,
\label{i.2}\endq
where $\rho_{\rm Min}(a^2;\ m_1,\ m_2)$ is easily explicitly computable simply by Fourier transform.

A similar explicit result has been recently obtained by the authors
for the de Sitter space-time \cite{bros3}. The derivation is considerably
more involved. This type of formula is of
interest in itself from the point of view of special-function
theory and also of group theory. In quantum field theory it allows
the computation
of the lifetime of a de Sitterian unstable particle at first order in perturbation theory: this was carried out in
\cite{bros1,bros2,bros3}.

Can the analogue of (\ref{i.2}) be explicitly obtained in the case of the AdS 
space-time? The general problem of constructing 
 the K\"all\'en-Lehmann representation for two-point functions of 
Anti-de Sitterian scalar fields was solved in \cite{dusedau} and a 
method of calculating the weight outlined there. Having such a 
representation is of course of importance  for  calculations in  
interacting Anti-de Sitter quantum field theories 
\cite{Dusedau:1985uf,Kabat:2011rz}. 

However to concretely derive an explicit expression for the weights in 
the quadratic case we study in this paper much additional effort is required.
With the notations of Sect. \ref{prelim}, in this paper we intend to 
establish that
\beq
W_{m+{d-1\over 2}}(z_1,\ z_2)\,W_{n+{d-1\over 2}}(z_1,\ z_2) =
\sum_l \rho(l;\ m,\ n) \,W_{l+{d-1\over 2}}(z_1,\ z_2)\ ,
\label{i.3}\endq
with an explicit determination of $\rho(l;\ m,\ n)$. Here $m$,
$n$ and $l$ take integer values. This will be done in the
following sections, as well as an extension to the case of
the universal cover of the AdS space-time.
\label{expan}
Equations (\ref{i.3}) and (\ref{s.10}) lead to
conjecture the following identity
\beq
D^{\lambda}_{m}(z) D^{\lambda}_{n}(z) =
\sum_l c_\lambda(m,n|l) D^{\lambda}_{l}(z)\ .
\label{p.1}\endq
in which we only suppose at first that $\Re( m+2\lambda) >0$ and
$\Re (n+2\lambda) >0$.

Formula (\ref{s.11}) shows that
$z^{n+2\lambda}D_n^\lambda(z)$ is holomorphic
and even in a neighborhood of $\infty$. It follows that in the rhs of
(\ref{p.1}), $l$ must take values of the form $l = m+n+2\lambda+2k$,
with integer $k \ge 0$. Inserting Eq. (\ref{s.11}) into
Eq.(\ref{p.1}) leads to yet another form of the conjectured identity:
\begin{align}
&F \left ({m+2\lambda\over 2},\ {m+2\lambda+1\over 2}\ ;\
m+\lambda+1\ ;\ u\right )
F \left ({n+2\lambda\over 2},\ {n+2\lambda+1\over 2}\ ;\
n+\lambda+1\ ;\ u\right ) =\cr
=& \sum_{k=0}^\infty b_\lambda(m,n|k)\,  u^k \,
F \left ({m+n+4\lambda+2k\over 2},\, {m+n+4\lambda+2k+1\over 2}\, ;\
m+n+3\lambda+1+2k\, ;\ u\right) .
\label{p.2}\end{align}
Here we have set $z^{-2} = u $ and adopted the definition
\beq
b_\lambda(m,n|k) =
\frac{ c_\lambda(m,n|m+n+2\lambda+2k)\ \alpha_\lambda(m+n+3\lambda+2k) }
{4^k \pi \alpha_\lambda(m + \lambda)\ \alpha_\lambda(n + \lambda)}\ .
\label{p.3}\endq
Using (\ref{s.12}) instead of (\ref{s.11}), or more directly
the identity (\ref{d.5}) of Appendix \ref{hypid}, we obtain the
equivalent (conjectured) identity
\begin{align}
&F(m+2\lambda,\ \lambda\ ;\ m+\lambda+1\ ;\ v)
F(n+2\lambda,\ \lambda\ ;\ n+\lambda+1\ ;\ v) =\cr
&= \sum_{k=0}^{\infty} b_\lambda(m,n|k)\,(4v)^k\,
F(m+n+4\lambda+2k,\ \lambda\ ;\ m+n+3\lambda+1+2k\, ;\ v) .
\label{p.4}\end{align}
If $u$ and $v$ are taken to be related by
\beq
u = z^{-2},\ \ \ z = \half(\zeta+\zeta^{-1}),\ \ \ v = \zeta^{-2},
\label{p.5}\endq
the series on the rhs of (\ref{p.2}) and (\ref{p.4}) are the same,
and the lhs are also the same.

For fixed $m$, $n$ and $\lambda$, identifying the power series
in $u$ which appear on both sides of (\ref{p.2}) allows an inductive
determination of the coefficients $b_\lambda(m,n|k)$. Identifying
the power series in $v$ which appear on both sides of (\ref{p.4})
leads to an equivalent algebraic problem.
This algebraic side of the problem will be discussed in the next section.

\section{The algebraic problem}
\label{alg}
It is useful to adopt as independent variables
$
x = m+2\lambda,\  y = n+2\lambda,\  \eta= 1-\lambda
$
instead of $m$, $n$, and $\lambda$, and to define
\beq
f_k(x,\ y,\ \eta) = 4^k b_\lambda(m,n|k)\ .
\label{a.2}\endq
Eq. (\ref{p.2}) then becomes
\begin{align}
\sum_{k=0}^\infty 4^{-k}\,f_k(x,\ y,\ \eta) u^k
F\left({x+y+2k\over 2},\ {x+y+2k+1\over 2}\ ;\
x+y+\eta+2k\ ;\ u \right ) =\cr
= F\left({x\over 2},\ {x+1\over 2}\ ;\ x+\eta\ ;\ u \right )
F\left({y\over 2},\ {y+1\over 2}\ ;\ y+\eta\ ;\ u \right )\ .
\label{a.3}
\end{align}
By using the Legendre duplication formula (\cite[1.3 (15) p.~5]{HTF1})  the relevant hypergeometric series simplifies as follows
\beq F\left (a,\ a+{1\over 2}\ ;\ c\ ;\ u \right ) 
=\sum_{p=0}^\infty \,\frac 1 {p!} {(2a)_{2p}\over (c)_p}\ \left(\frac u 4 \right)^p.\label{a.5}\endq
and equating the coefficients of $u^r$ on both sides of Eq. (\ref{a.3}) gives
\begin{align}
& \sum_{k=0}^{r} \,f_k(x,\ y,\ \eta)\,
{(x+y+2k)_{2(r-k)}\over
(x+y+\eta+2k)_{r-k}\,(r-k)!}
=\sum_{p=0}^{r} {(x)_{2p}\,(y)_{2{r-2p}}\over
(x+\eta)_p\,p!\,(y+\eta)_{r-p}\,{(r-p)}!}  .
\label{a.6}\end{align}
Note the convolution structure of the rhs;
the coefficients there can be seen to be a one-parameter deformation of 
the binomial coefficient $\binom{x+2p-1}{p}$.
\beq
{\binom{x+2p-1}{p}}_\eta= {(x)_{2p}\over
(x+\eta)_p\,p!} = \frac{\Gamma(x+2p)}{\Gamma(x+\eta+p)\Gamma(p+1)}
\frac{\Gamma(x+\eta)}{\Gamma(x)}.
\endq
The system (\ref{a.6}) is suitable for an iterative solution of the problem.

Setting $r=0$  gives $f_0(x,\ y,\ \eta) = 1$. For any $r>0$, the coefficient
of $f_r(x,\ y,\ \eta)$ in Eq. (\ref{a.6}) is 1 so that (\ref{a.6})
provides an expression
of $f_r(x,\ y,\ \eta)$ in terms of all the $f_k(x,\ y,\ \eta)$ with $k<r$.

It is clear by induction that all
$f_r(x,\ y,\ \eta)$ are rational functions of the variables $x$,
$y$, and $\eta$, the degrees of numerator and denominator depending
on $r$. An equivalent form of the system of equations (\ref{a.6})
is obtained by defining
\beq
a_r(x,\ y,\ \eta) = r!f_r(x,\ y,\ \eta)(x+\eta)_r(y+\eta)_r
(x+y+\eta)_{2r}\ .
\label{a.7}\endq
Then (\ref{a.6}) gives
\begin{align}
&\sum_{k=0}^{r}
a_k(x,\ y,\ \eta)\,\binom{r}{k}\, (x+y+2k)_{2(r-k)}  (x+\eta+k)_{r-k}\,(y+\eta+k)_{r-k}
(x+y+\eta+k+r)_{r-k}  = \cr
&= (x+y+\eta)_{2r}\,
\sum_{p=0}^{r} \binom{r}{p}\,(x)_{2p}\,(x+\eta+p)_{r-p}\,
(y)_{2r-2p}\,(y+\eta+r-p)_p\ .
\label{a.8}
\end{align}
The case $r=0$ gives $a_0(x,\ y,\ \eta) = 1$. Since
the binomial coefficient is an integer,
for $r >0$ the function $a_r(x,\ y,\ \eta)$ is seen
by induction to be a polynomial with integer coefficients
in the variables $x$, $y$, and $\eta$,
with degree $\le 5r$, degree in $x$ or $y$ $\le 4r$,
and degree in $\eta$ $\le 3r$.

With the same change of variables Eq. (\ref{p.4})
becomes
\begin{align}
&F(x,\ 1-\eta\ ;\ x+\eta\ ;\ v)
F(y,\ 1-\eta\ ;\ y+\eta\ ;\ v) =\cr
&= \sum_{k=0}^{\infty} \,f_k(x,\ y,\ \eta) v^k
F(x+y+2k,\ 1-\eta\ ;\ x+y+\eta+2k\ ;\ v)\ .
\label{y.0}\end{align}
Equating the terms in $v^r$ on both sides of (\ref{y.0}) shows that
for every integer $r \ge 0$,
\begin{align}
&\sum_{k=0}^{r} \,f_k(x,\ y,\ \eta)\,
{(x+y+2k)_{r-k}\,(1-\eta)_{r-k}\over
(x+y+\eta +2k)_{r-k} (r-k)!} = \sum_{p,\ q \ge 0,\ p+ q = r} {(x)_p(1-\eta)_p(y)_q(1-\eta)_q\over
(x+\eta)_p(y+\eta)_q\,p!q!}
\label{y.1}\end{align}
and, with the same definition of $a_k(x,\ y,\ \eta)$ as in (\ref{a.7}),
\begin{align}
&\sum_{k=0}^{r} \binom{r}{k}\,a_k(x,\ y,\ \eta)\times\cr
&\times (x+\eta+k)_{r-k}(y+\eta+k)_{r-k} (x+y+\eta+r+k)_{r-k}
(x+y+2k)_{r-k} (1-\eta)_{r-k} =\cr
&=(x+y+\eta)_{2r}\sum_{p,\ q \ge 0,\ p+ q = r} \binom{r}{p}\,
(x)_p(1-\eta)_p(x+\eta+p)_q
(y)_q(1-\eta)_q(y+\eta+q)_p\ .
\label{y.2}\end{align}
Of course the systems
(\ref{a.6}), (\ref{a.8}), (\ref{y.1}) and (\ref{y.2}) are
all equivalent. Note that from the form (\ref{y.2}) it can be seen
by induction that $a_r(x,\ y,\ \eta)$ is of degree $\le 3r$ in
$x$ and of degree $\le 3r$ in $y$.

From now on, $f_r(x,\ y,\ \eta)$ and $a_r(x,\ y,\ \eta)$ will denote
the solutions of the systems (\ref{a.6}) or (\ref{y.1})
and (\ref{a.8}) or (\ref{y.2}), respectively
(they are, of course, related by (\ref{a.7})), and
$c_\lambda(m,n|l)$ will denote the quantity obtained from this
solution by retracing through Eqs. (\ref{a.2})
and (\ref{p.3}).

The following theorem will be proved:

\begin{theorem}
\label{main}
The solution of the system (\ref{a.6}) or (\ref{y.1}) is explicitly given by
\beq
f_k(x,\ y,\ \eta) = \wt f_k(x,\ y,\ \eta) \bydef
{(x)_k(y)_k (x+y+2\eta-1+k)_k(1-\eta)_k\over
(x+\eta)_k (y+\eta)_k (x+y+\eta-1+k)_k\,k!}\ .
\label{a.9}\endq
Equivalently, the solution of (\ref{a.8}) or (\ref{y.2})
is explicitly given by
$a_k(x,\ y,\ \eta) = \wt a_k(x,\ y,\ \eta)$, with
$\wt a_0(x,\ y,\ \eta) = 1$ and, for $k \ge 1$,
\beq
\wt a_k(x,\ y,\ \eta) \bydef
(x+y+\eta+2k-1)(x)_k(y)_k
(x+y+\eta)_{k-1} (x+y+2\eta -1+k)_k (1-\eta)_k\ .
\label{a.10}\endq
Finally redefining $c_\lambda(m,n|l)$ by inverting Eqs. (\ref{a.2})
and (\ref{p.3}), we get, for $l=m+n+2\lambda +2k$, $k$ a non-negative
integer,
\beq
c_\lambda(m,n|l) =
\frac{\pi {\alpha_\lambda\left(\frac{l+m-n}{2}\right)}
 {\alpha_\lambda\left(\frac{l-m+n}{2}\right)}
 {\alpha_\lambda\left(\frac{l+m+n}{2}+\lambda\right)}
{\alpha_\lambda\left(\frac{l-m-n}{2}-\lambda\right)}}{ \alpha_\lambda(l)
\alpha_\lambda\left(l+\lambda \right)}
\label{a.11}\endq
\end{theorem}
In this statement, (\ref{a.10}) is an identity between polynomials
in $x$, $y$ and $\eta$, but (\ref{a.9}) must be understood as
an identity between rational functions, and (\ref{a.11}) as an
identity between meromorphic functions.

It is remarkable that the polynomial $\wt a_k(x,\ y,\ \eta)$ is
completely factorized into a product of polynomials of the first
degree (with integer coefficients) in all variables. This gives
the identities (\ref{a.8}) and (\ref{y.2}) the appearance of a small
algebraic miracle.

We do not have, at the moment, a purely algebraic proof of this
theorem. Instead it will be proved by a roundabout analytic method.
We will use the following remark:

\begin{remark}[Algebraic continuation]
\label{poly}\rm
Let $S(x_1,\ldots,\ x_N)$ be a complex polynomial in $N$ variables
of degree $d_j$ in $x_j$ for all $j$. Suppose that $S$ vanishes
on $\AA_1 \times \ldots \times \AA_N$ where, for each $j$, $\AA_j \subset \bC$
has more than $d_j$ distinct elements. Then, as it is easy to see
by induction on $N$, $S$ is identically 0. This is in particular true
if all the $\AA_j$ are infinite.

\end{remark}

Let us now assume that, for some fixed $k > 0$,
the statement of the theorem, in the form
(\ref{a.11}), has been proved under the following
assumptions: $m$, $n$, $r = \half -\lambda$ are non-negative integers and
\beq
m -2r \ge 0,\ \ \ n -2r \ge 0,\ \ \ r \ge 1\ .
\label{a.12}\endq
(It follows that $l=m+n+2\lambda+2k$ is an integer verifying
$l-m > 0$, $l-n >0$, $l-2r > 0$.)
This is equivalent to having proved (\ref{a.9}) and (\ref{a.10})
(for this value of $k$) under the conditions
that $x$, $y$, $\eta-\half$ are integers satisfying
\beq
\eta-\half \ge 1,\ \ \ x-1 \ge 0,\ \ \ y-1 \ge 0\ .
\label{a.13}\endq
Then, by applying Remark \ref{poly} to
$a_k(x,\ y,\ \eta) - \wt a_k(x,\ y,\ \eta)$, considered as a polynomial in
$x,\ y,\ \eta$, we conclude that this polynomial is identically 0,
i.e. that, for that value of $k$, the theorem holds for all values
of $x,\ y,\ \eta$. This will be done, and Theorem \ref{main} will
be proved, in Sect. \ref{ana}.

\section{Checking the conjecture}
\subsection{Computer proofs at fixed $r$}

According to (\ref{a.8})
proving Theorem \ref{main} is equivalent to proving that,
for every integer $r >0$, the two
polynomials with integer coefficients
\begin{align}
L_r(x,\ y,\ \eta) &=
\sum_{k=0}^{r} {r!\over k!(r-k)!}
\wt a_k(x,\ y,\ \eta)\, (x+y+2k)_{2(r-k)}\,
(x+\eta+k)_{r-k}\,(y+\eta+k)_{r-k} \times\cr
&\hskip 2 cm \times (x+y+\eta+k+r)_{r-k}\ ,
\label{a.21}\end{align}
and
\beq
R_r(x,\ y,\ \eta) =
(x+y+\eta)_{2r}\,
\sum_{p,\ q \ge 0,\ p+ q = r} {r!\over p! q!}\,(x)_{2p}\,(x+\eta+p)_q\,
(y)_{2q}\,(y+\eta+q)_p
\label{a.22}\endq
coincide. Here $\wt a_0(x,\ y,\ \eta) =1$, and, for $k \ge 1$,
$\wt a_k(x,\ y,\ \eta)$ is given by (\ref{a.10}).
Equivalently, by (\ref{y.2}),
\begin{align}
L'_r(x,\ y,\ \eta) &=
\sum_{k=0}^{r} {r!\over k!(r-k)!}
\wt a_k(x,\ y,\ \eta)\, (x+\eta+k)_{r-k}\,(y+\eta+k)_{r-k}(1-\eta)_{r-k}
\times\cr
&\hskip 2 cm \times (x+y+\eta+k+r)_{r-k}(x+y+2k)_{r-k}
\label{a.23}\end{align}
and
\beq
R'_r(x,\ y,\ \eta) = (x+y+\eta)_{2r}\,
\sum_{p,\ q \ge 0,\ p+ q = r} {r!\over p! q!}\,
(x)_p(1-\eta)_p(x+\eta+p)_q
(y)_q(1-\eta)_q(y+\eta+q)_p
\label{a.24}\endq
must coincide.
Let $r \ge 1$ be fixed. $L_r$ and $R_r$ have degree $\le 4r$ in $x$,
degree $\le 4r$ in $y$, and degree $\le 3r$ in $\eta$. By Remark \ref{poly}
it suffices to check that $L_r(x,\ y,\ \eta)$ and $R_r(x,\ y,\ \eta)$
take the same values when $(x,\ y,\ \eta)$ runs over a set of
the form $\AA_1 \times \AA_2 \times \AA_3$, where the $\AA_j$
are finite sets with (at least) $4r+1$, $4r+1$ and $3r+1$
elements, respectively. For example we can take
$\AA_1 = \AA_2 = \bZ\cap [-4r,\ 0]$, $\AA_3 = \bZ\cap [0,\ 3r]$.
It is therefore possible to write a program to prove
the conjecture for any fixed $r$ using some form of
arbitrarily large integer, such as GNU's GMP's {\tt mpz\_t}
or Java's BigInteger.
A few easy remarks allow to omit checking for some of these values:
the symmetry in $x$ and $y$, the identities
\beq
L_r(0,\ y,\ \eta) = R_r(0,\ y,\ \eta) = (y)_{2r} (y+\eta)_{2r}(\eta)_r\ ,
\ \ L_r(x,\ 0,\ \eta) = R_r(x,\ 0,\ \eta) = (x)_{2r} (x+\eta)_{2r}(\eta)_r
\label{a.24.1}\endq
and the fact that $L_r(x,\ y,\ \eta) = R_r(x,\ y,\ \eta) = 0$ if
$x$ and $y$ are negative integers such that $2r+x+y > 0$. It is
thus sufficient to check only the integer values of $x$, $y$ and $\eta$
such that
\beq
-4r \le x \le -r,\ \ \ x \le y < 0,\ \ \ y \le -x-2r,\
\ \ 0\le \eta \le 3r\ .
\label{a.24.2}\endq
Similar remarks apply to the second form of the problem,
$L'_r(x,\ y,\ \eta) = R'_r(x,\ y,\ \eta)$, which is even more favorable
as we can take
$\AA_1 = \AA_2 = \bZ\cap [-3r,\ 0]$, $\AA_3 = \bZ\cap [0,\ 4r]$,
and use the easily checked identity
\beq
L'_r(x,\ y,\ 0) = R'_r(x,\ y,\ 0) = (r+1)! (x)_r (y)_r (x+y)_{2r}\ .
\label{a.24.3}\endq
Note also that if $\eta$ is an integer and $0\le \eta-1 < r/2$,
then $L'_r(x,\ y,\ \eta) = R'_r(x,\ y,\ \eta) = 0$.
It is therefore sufficient to check only the
integer values of $x$, $y$ and $\eta$ such that
\beq
-3r \le x \le -{r\over 2},\ \ \ x \le y < 0,\ \ \ y \le -x-r,\ \ \
1+{r\over 2} \le \eta \le 4r\ .
\label{a.24.4}\endq
We have used such programs
to prove the theorem for $r= 1,\ldots,\ 51$,
$r = 101,\ 151,\ 171,\ 201,\ 250,\ 301$\footnote{
Computing time appears to increase very roughly like $r^4$.
On a machine with two INTEL XEON E5504 processors at 2GH, each with
4 cores, $r=301$ (first form) took approximately 55 days.}.
Of course these proofs for selected values of $r$ do not constitute a
computer-proof of Theorem \ref{main}, but they at least serve as
a check on the calculations involved in the actual proof (Sect. \ref{ana}).

\subsection{The three dimensional case $d=3$}
\label{dim3}
The three-dimensional case $\lambda = \frac {d-1} 2 = 1$ (i.e. $\eta = 0$)
allows for a simple verification of the different forms of the conjecture.

Let us verify at first  formula (\ref{p.1}).
Here the conjectured coefficients do not depend on $n,m$ and $l$ and have all the same value
\begin{eqnarray}
c_{1}(m,n|l)&=& \pi.
\end{eqnarray}
The three-dimensional Gegenbauer function (\ref{s.11}) is most simply written by using the variable $z=\cosh t$:
\begin{eqnarray}
 {D}^{1}_n(\cosh t ) &=&   \frac \pi 2  \frac {e^{-(n+1)t}} {\sinh t }
\end{eqnarray}
Suppose that  $\Re t>0$; it follows that
\begin{eqnarray}
{D}^{1}_m(\cosh t ){D}^{1}_n(\cosh t ) &=&  \frac {\pi^2} 4 \,   \frac {e^{-(m+n+2)t}} {\sinh^2 t } =
  \frac {\pi^2} 2   \frac {e^{-t}} {1 - e^{-2t}} \frac {e^{-(m+n+2)t}} {\sinh t } =
 \cr &=&
  \frac {\pi^2} 2  \sum_{k=0}^{\infty} \frac {e^{-(m+n+3+ 2k)t}} {\sinh t }  =  \pi   \sum_{k=0}^{\infty} {D}^{1}_{m+n+ 2+2k}(\cosh t )
\end{eqnarray}
and the conjectured formula (\ref{p.1}) is readily verified.

It is equally simple to verify directly the conjecture in the form given by
Eq. (\ref{p.2}); the conjectured coefficients are  $b_{1}(m,n|k)= 4^{-k}$.
Since
\begin{eqnarray}
F\left(a,a+ \frac 12; 2a; u\right)  = \frac {1}{\sqrt{1-u}} \left(\frac {2}{1+\sqrt{1-u}}\right)^{2a-1}
\end{eqnarray}
it follows that
\begin{eqnarray}
&& \sum_k   \left(\frac u 4\right)^k F\left(\frac{m+n+4 +2k}{2},\frac{m+n+5+2k}{2} ;m+n+4+2k;u\right) = \cr
 && = \frac {1}{\sqrt{1-u}}  \left(\frac {2}{1+\sqrt{1-u}}\right)^{m+n+3}
\sum_k   \left(\frac {\sqrt u}{1+\sqrt{1-u}}\right)^{2k}
=\frac {1}{{1-u}}  \left(\frac {2}{1+\sqrt{1-u}}\right)^{m+n+2}
= \cr && = F\left(\frac{m+2}{2},\frac{m+3}{2};m+2;u\right)
F\left(\frac{n+2}{2},\frac{n+3}{2};n+2;u\right) \label{ppp}
\end{eqnarray}
The second form of the algebraic problem is also immediately
verified by Eq. (\ref{a.24.3}).
Verifying the conjecture order by order as in Eq. (\ref{a.6})  is a little trickier already in this elementary case.
At order $r$ the validity of the conjecture amounts in $d=3$ to the following identity:
\begin{align}
\sum_{s=0}^{r} \binom{x+y+2r-1}{r-s}=
\sum_{p=0}^{ r} \binom {x+2p-1}{p} \binom{y+2r-2p-1}{r-p}  .
\label{a3d.6}
\end{align}
A formula closely similar to this one has been proven in
\cite[Eq. (5.13-14)]{bouttier}.
A direct bijective proof\footnote{This proof has been communicated to us by J. Bouttier and E. Guitter. We thank them for discussions on this point} is as follows. The lhs counts random
walks of length $x+y+2r-1$, starting at height 0 and ending at height
$\geq x+y-1$. The same set of walks is counted differently at the rhs.
First, one performs a last-passage decomposition at height
$x-1$; the length is of the form $x-1+2p$ with $0 \leq p \leq
r$. This yields immediately
\begin{equation}
\sum_{k=0}^r \binom{x+y+2r-1}{k} = \sum_{p=0}^r \binom{x+2p-1}{p} B(y,r-p)
\end{equation}
where $B(y,q)$ is the number of \emph{positive} walks of length
$y-1+2q$ starting at height 0 and ending at height $\geq y-1$. It
remains to show that $B(y,q)=\binom{y-1+2q}{q}$. Consider such a walk
ending at height $y-1+2j$, and consider the last passages at heights
$0,1,\ldots,j-1$: ``flip'' the corresponding steps (from up to
down). This defines a bijection with walks of length $y-1+2q$ starting
at height 0 and ending at $y-1$ whose minimal height is $-j$ (the
flipped steps become the first passages at heights
$-1,-2,\cdots,-j$). Summing over $j$ the result follows.
\subsection{Remark}
The validity of the relation (\ref{a3d.6})
is already guaranteed by the previous Eq. (\ref{ppp}).
Similarly, the proof of the full conjecture will imply the validity of
the following one parameter deformation  of Eq. (\ref{a3d.6})
\begin{align}
&\sum_{k=0}^{r} \,\frac{(x)_k}{(x+\eta)_k} \,\frac{(y)_k}{(y+\eta)_k}
\frac{(x+y+2\eta-1+k)_k (x+y+2k)_{r-k}}{(x+y+\eta-1+k)_k (x+y+\eta +2k)_{r-k}}
{(1-\eta)_k (1-\eta)_{r-k}\over
 k!(r-k)!} {\binom{x+y+2r-1}{k}}_\eta = \cr   
&= \sum_{p=0}^{r} {\binom{x}{p}}_\eta {\binom{y}{r-p}}_\eta
\label{yyy1}
\end{align}
At the moment we do not know if any combinatorial interpretation of this
generalization of Eq. (\ref{a3d.6}) does exist.

\section{An analytic version of the problem}
\label{ana}

In this section $\lambda$ will be of the form
$\lambda = \half - r$, and $r$,
$m$, and $n$ will be integers such that
\beq
r \ge 1,\ \ \ \ m-2r \ge 0,\ \ \ \ n-2r \ge 0\ .
\label{b.0}\endq
Under these conditions the function
\beq
F(z) = (z^2-1)^{\lambda - \half}\,D^{\lambda}_{m}(z) D^{\lambda}_{n}(z)
\label{b.0.1}\endq
is holomorphic on $\wh \Delta_1$ and
satisfies the hypotheses of Theorem \ref{dexp},
stated in Appendix \ref{jac} with
$N= m+n+2-2r$, $N-1 = m+n+1-2r \ge 2r+1$.
The theorem then asserts that
\beq
(z^2-1)^{\lambda - \half}\,D^{\lambda}_{m}(z) D^{\lambda}_{n}(z) =
(z^2-1)^{\lambda - \half}\,\sum_{l= m+n+2\lambda}^\infty
\wh c_\lambda(m,n|l) D^{\lambda}_{l}(z)
\label{b.1}\endq
holds with uniform convergence on every compact subset of
$\wh \Delta_1$, with
\beq
\wh c_\lambda(m,n|l) = {2^{2\lambda-1}(l+\lambda)
\Gamma(l+1)\Gamma(\lambda)^2\over i\pi^2 \Gamma(l+2\lambda)}
\int_{E(L)} C_l^\lambda(z)\,(z^2-1)^{\lambda - \half}\,
D^{\lambda}_{m}(z) D^{\lambda}_{n}(z)\,dz\ .
\label{b.2}\endq
Here $E(L)$ is the ellipse defined in (\ref{s.14}) for any $L>1$,
but it can, of course, be replaced by any continuous contour
homotopic to it in $\Delta_1$. The index $l$ takes integer values.
It follows from $C_l^\lambda(z) = (-1)^l C_l^\lambda(-z)$
that $\wh c_\lambda(m,n|l)$ vanishes unless $l=m+n+2\lambda+2k$
with an integer $k \ge 0$.
Since the Laurent coefficients of both sides of (\ref{b.1})
can be obtained by
Cauchy integrals on some circle centered at 0 with radius $R>1$,
and the expansion is uniform on this circle, the coefficients
$\wh c_\lambda(m,n|l)$ can be obtained by identifying
these Laurent series. Hence $\wh c_\lambda(m,n|l) = c_\lambda(m,n|l)$,
where $c_\lambda(m,n|l)$ is the solution of the algebraic problem
considered in sect. \ref{alg}, and (\ref{b.2}) gives a new expression
for this solution under the special conditions we have imposed.

Since the function $F$ is holomorphic in $\Delta_1$ and has
tempered behavior, we find by contour deformation
\beq
\int_{E(L)} C_n^{\half-r}(z)\,F(z)\,dz =
-\int {\rm disc\,}\left [ C_n^{\half-r}(x)\, F(x)\right ]\,dx \ .
\label{b.4}\endq
Therefore
\begin{align}
c_{\half-r}(m,n|l) &= {-2^{-2r}(l+\half-r)
\Gamma(l+1)\Gamma(\half-r)^2\over i\pi^2 \Gamma(l+1-2r)}
\int {\rm disc\,}\left [(x^2-1)^{-r}\,C_l^{\half-r}(x)\,
D^{\half-r}_{m}(x) D^{\half-r}_{n}(x)\right ]\,dx
\label{b.6}\\
&=  {-2^{-2r}(l+\half-r)
\Gamma(l+1)\Gamma(\half-r)^2\over i\pi^2 \Gamma(l+1-2r)}
\int (x^2-1)^{-r}\,C_l^{\half-r}(x)\,{\rm disc\,}\left [
D^{\half-r}_{m}(x) D^{\half-r}_{n}(x)\right ]\,dx\ .
\label{b.7}\end{align}
The last equality holds because in the case $r \ge 1$, $l \ge 2r$,
which we are considering, $(x^2-1)^{-r}\,C_l^{\half-r}(x)$ is
a polynomial. Under our assumptions, $D^{\half-r}_{m}$ and $D^{\half-r}_{n}$
have continuous boundary values on the real axis (see  Appendix \ref{jac},
subsect. \ref{jacspec}). Thus the integrand of (\ref{b.6})
or (\ref{b.7})is a continuous function on $\bR$ with support in $[-1,\ 1]$.

Since the boundary
values of $D^{\half-r}_{k}$ are continuous, it suffices to study
${\rm disc\,}\left [D^{\half-r}_{m}(x) D^{\half-r}_{n}(x)\right ]$
on the open interval $(-1,\ 1)$. It is given by
\beq
{\rm disc\,}\left [
D^{\half-r}_{m}(x) D^{\half-r}_{n}(x)\right ] =
[{\rm disc\,}D^{\half-r}_{m}(x)]\,\D^{\half-r}_{n}(x) +
\D^{\half-r}_{m}(x)\,[{\rm disc\,}D^{\half-r}_{n}(x)]\ .
\label{b.8}\endq
Here $\D^{\half-r}_{k}(x)$ is defined as
$\half D^{\half-r}_{k}(x+i0)+ \half D^{\half-r}_{k}(x-i0)$ and is discussed
in Appendix \ref{jac}, as well as ${\rm disc\,}D^{\half-r}_{n}(x)$.
The integrand of (\ref{b.7}) is equal to
\beq
I = I_{m,n} + I_{n,m},
\label{b.9}\endq
\begin{align}
I_{m,n} &= {(-i)\pi^{3/2} (-2)^r \Gamma(n-2r+1)\over
\Gamma(n+1)\Gamma(\half-r)}\,\theta(1+x)\theta(1-x) \times\cr
&\times (1-x^2)^{-r/2}\,C_l^{\half-r}(x)\,C_m^{\half-r}(x)\,\Q_{n-r}^r(x)\ .
\label{b.10}\end{align}
Since $l > 2r$, $I_{m,n}$ is of the form
\beq
I_{m,n} = h(x)\,\Q_{n-r}^r(x)\ ,\ \ \ \
h(x) = \theta(1+x)\theta(1-x)(1-x^2)^{3r/2}\vhi(x),
\label{b.11}\endq
where $\vhi$ is a polynomial. Thus
\begin{align}
& (1-x^2)^{r/2} h(x) = \theta(1+x)\theta(1-x)(1-x^2)^{2r}\vhi(x),
\label{b.11.1}\\
& \left ({d\over dx}\right )^r \left [h(x)(1-x^2)^{r/2}\right ] =
\theta(1+x)\theta(1-x) \left ({d\over dx}\right )^r \left [
(1-x^2)^{2r}\vhi(x) \right ]\ .
\label{b.11.2}\end{align}
The last function (\ref{b.11.2}) is continuous on $\bR$ with support
in $[-1,\ 1]$, thus belongs to $L^2(\bR)$.
For reasons explained in Appendix \ref{jac}
(subsect. \ref{qrpr}),
it is legitimate to make the substitution
\beq
\Q_{n-r}^r(x) = \sum_{k=2r}^\infty a_r(k,\ n) \P_{k-r}^r(x)\ ,
\label{b.12}\endq
where
\beq
a_r(k,\ n) = a(k-r,\ n-r),
\label{b.13}\endq
and
\beq
a(k,\ n) = \left \{
\begin{array}{ll}
\displaystyle {1\over k-n} + {1 \over k+n+1} & {\rm if\ } k-n\ {\rm is\ odd}\\
\hbox to 1cm{\hfill}\\
0 & {\rm if\ } k-n\ {\rm is\ even}
\end{array} \right .
\label{b.14}\endq
(These $a_r$ have nothing to do with those used in sect. \ref{alg}).
Using the identity (see \cite[3.15.1 (4), p.~175]{HTF1} and (\ref{u.9.3}))
\beq
\P_{k-r}^r(x) =
{2^{-r}\pi^{-1/2} \Gamma(\half-r)\Gamma(k+1)\over \Gamma(k-2r+1)}
(1-x^2)^{-r/2}C_k^{\half-r}(x)\ ,
\label{b.15}\endq
we obtain
\beq
c_{\half-r}(m,n|l) = R(m,\ n|l) + R(n,\ m|l),
\label{b.16}\endq
\begin{align}
&R(m,\ n|l) = {(-4)^{-r}(l+\half-r)
\Gamma(l+1)\Gamma(\half-r)^2\over \pi \Gamma(l+1-2r)}\times \cr
&\sum_{k=2r}^\infty {\Gamma(n-2r+1)\Gamma(k+1)\over
\Gamma(k-2r+1)\Gamma(n+1)} a_r(k,\ n)\,H(r\ ;\ m,\ k,\ l)\ ,
\label{b.17}\end{align}
with
\beq
H(r\ ;\ n_1,\ n_2,\ n_3) =
\int_{-1}^1
(1-x^2)^{-r}\,C_{n_1}^{\half-r}(x)\,C_{n_2}^{\half-r}(x)\,C_{n_3}^{\half-r}(x)\,dx\ .
\label{b.18}\endq
An explicit expression for $H(r\ ;\ n_1,\ n_2,\ n_3)$ has been given
by Hs\"u (\cite{Hs}) for any triple of integers $n_1,\ n_2,\ n_3$ and
any complex $r$ with $\Re r < 1$. (See also the very interesting
discussions in \cite{A1} and \cite{A2} for the connection with a
theorem of Dougall.)
In Appendix \ref{hsu}, this result
is shown also to hold for all {\it integer} values of $r >0$ such that
$n_j \ge 2r$. Recall that for such values, each of the polynomials
$C_{n_j-r}^r(x)$ is divisible by $(1-x^2)^{r}$. It is found that
$H(r\ ;\ n_1,\ n_2,\ n_3)$ is equal to 0 unless $n_j \le n_k+n_l$
for any permutation $(j,\ k,\ l)$ of $(1,\ 2,\ 3)$, and
$2s = n_1+n_2+n_3$ is an even integer. Otherwise,
\begin{align}
&H(r\ ;\ n_1,\ n_2,\ n_3) = \cr
&2^r\pi {\Gamma(s+1-2r) \over \Gamma \left (s+{3\over 2}-r \right )
\Gamma(\half -r)}
{\Gamma(s-n_1+\half -r) \over \Gamma(s-n_1+1) \Gamma(\half -r)}
{\Gamma(s-n_2+\half -r) \over \Gamma(s-n_2+1) \Gamma(\half -r)}
{\Gamma(s-n_3+\half -r) \over \Gamma(s-n_3+1) \Gamma(\half -r)}\ .
\label{b.19}\end{align}
An important consequence of this is that the summation in (\ref{b.17})
extends only over values of $k$ such that $l-m \le k \le l+m$.

\newpage
\subsection{Evaluation}
We will now proceed to the actual evaluation of $R(m,\ n|l)$.
To this end we consider the meromorphic function $x \mapsto s(x)$
defined by
\beq
s(x) =
\frac {\Gamma\left(\frac{l+m-x+1-2r}{2}\right)}
{\Gamma\left(\frac{l+m-x+2}{2}\right)}
\frac {\Gamma\left(\frac{l+m+x+2-4r}{2}\right)}
{\Gamma\left(\frac{l+m+x+3-2r}{2}\right)}
\frac {\Gamma\left(\frac{l-m+x+1-2r}{2}\right)}
{\Gamma\left(\frac{l-m+x+2}{2}\right)}
\frac {\Gamma\left(\frac{l-m-x}{2}\right)}
{\Gamma\left(\frac{l-m-x+1+2r}{2}\right)}
\frac {\Gamma\left(x+1\right)}{\Gamma\left(x-2r+1\right)}
\label{q.1}\endq
where we continue to suppose $l \ge m+n+2\lambda$, $\lambda = \half-r$,
$m$, $n$ and $r$ integers, $r \ge 1$, $m-2r \ge 0$,  $n-2r \ge 0$,
which imply $l-m >0$.
$s(x)$ is the ratio of two polynomials,
\beq
s(x) = {p(x)\over q(x)}\ ,
\label{q.2}\endq
where
\begin{align}
p(x) &= \frac{\Gamma\left(\frac{l+m+1-x-2r}{2}\right)}
{\Gamma\left(\frac{l-m+1-x+2r}{2}\right)}
\ \frac{\Gamma\left(\frac{l+m +x+2-4r}{2}\right)}
{\Gamma\left(\frac{l-m+x+2}{2}\right)}
\frac {\Gamma\left(x+1\right)}{\Gamma\left(x-2r+1\right)} =\cr
&= \left(\frac{l-m+1-x+2r}{2} \right)_{m-2r}
\left(\frac{l-m+x+2}{2}\right)_{m-2r} (x-2r+1)_{2r}\ ,
\label{q.3}\\
q(x) &= \frac {\Gamma\left(\frac{l+m+2-x}{2}\right)}
{\Gamma\left(\frac{l-m-x}{2}\right)}\ \
\frac {\Gamma\left(\frac{l+m+x+3-2r}{2}\right)}
{\Gamma\left(\frac{l-m+x+1-2r}{2}\right)}
=\left(\frac{l-m-x}{2} \right)_{m+1}
\left(\frac{l-m+x+1-2r}{2} \right)_{m+1}\ .
\label{q.4}\end{align}
The degree of $p(x)$ is $2m-2r$ while the degree of $q(x)$ is $2m +2$.
The symmetry $s(x) = s(-x+2r-1)$ implies that $s(x)$ admits a
partial fraction decomposition of the following form:
\beq
s(x) = \sum_{k=l-m,\, k\geq 2r}^{l+m} \sigma(k,l,m)
\left(\frac 1 {k-x} + \frac{1}{k+x-2r+1}\right)\ ,
\label{q.5}\endq
where the sum runs over the zeros of the polynomial
$\left(\frac{l-m-x}{2} \right)_{m+1}$ namely
\beq
k = l-m,\ l-m+2,\ \ldots,\ l+m\ .
\label{q.6}\endq
The coefficient $\sigma(k,l,m)$ may be computed by taking the residue
\beq
\sigma(k,l,m) =  \lim_{x\to k} (k-x) s(x)\ .
\label{q.7}\endq
Since
\beq
{\Gamma\left({-k+\epsilon}\right)}\thicksim
\frac{(-1)^{k}}{\Gamma\left(k+1\right)}\frac{1}{\epsilon}\ ,
\label{q.8}\endq
we find
\begin{align}
&\sigma(k,l,m)= \lim_{x\to k} (k-x) s(x) = \cr
&=\frac{2(-1)^{\frac{l-m-k}2}}{\Gamma\left(\frac{m+k-l+2}{2}\right)}
\frac {\Gamma\left(\frac{l+m-k+1-2r}{2}\right)}
{\Gamma\left(\frac{l+m-k+2}{2}\right)}
\frac {\Gamma\left(\frac{l+m+k+2-4r}{2}\right)}
{\Gamma\left(\frac{l+m+k+3-2r}{2}\right)}
\frac {\Gamma\left(\frac{l-m+k+1-2r}{2}\right)}
{\Gamma\left(\frac{l-m+k+2}{2}\right)}
\frac {1}{\Gamma\left(\frac{l-m-k+1+2r}{2}\right)}
\frac {\Gamma\left(k+1\right)}{\Gamma\left(k-2r+1\right)}= \cr
&=
\frac {2} {\pi}(-1)^{r}
\frac{\Gamma\left(\frac{m+k-l+1-2r}{2}\right)}
{\Gamma\left(\frac{m+k-l+2}{2}\right)}
\frac {\Gamma\left(\frac{l+m-k+1-2r}{2}\right)}
{\Gamma\left(\frac{l+m-k+2}{2}\right)}
\frac {\Gamma\left(\frac{l+m+k+2-4r}{2}\right)}
{\Gamma\left(\frac{l+m+k+3-2r}{2}\right)}
\frac {\Gamma\left(\frac{l-m+k+1-2r}{2}\right)}
{\Gamma\left(\frac{l-m+k+2}{2}\right)}
\frac {\Gamma\left(k+1\right)}{\Gamma\left(k-2r+1\right)}= \cr
&=
\frac {1} {2^{2r-1}\pi^2}(-1)^{r} \left[\Gamma\left(\frac 12 -r\right)\right]^4
\frac {\Gamma\left(k+1\right)}{\Gamma\left(k-2r+1\right)} H(m,k,l)\ .
\label{q.9}\end{align}
Therefore
\begin{align}
&R(m,n|l) = \cr
&=
\frac{2^{ -2 r }\cos r\pi \, [\Gamma(\frac 12-r)]^2 \left(l-r+\frac 12\right)}
{[\Gamma(\frac 12 )]^2} \frac{\Gamma(l+1)}{\Gamma(l-2r +1) }
\sum_{k=2r}^{l+m} \frac{\Gamma(n-2r+1) \Gamma(k+1)}{\Gamma(n+1)\Gamma(k-2r+1) }
 a_r(k,\ n) H(m,\ k,\ l) =\cr
&=
2^{2r-1}\pi^2 (-1)^{r} \frac 1{\left[\Gamma\left(\frac 12 -r\right)\right]^4}
\frac{2^{ -2 r }\cos r\pi \, [\Gamma(\frac 12-r)]^2 \left(l-r+\frac 12\right)}
{[\Gamma(\frac 12 )]^2}
\frac{\Gamma(l+1)}{\Gamma(l-2r +1) } \frac{\Gamma(n-2r+1)}{\Gamma(n+1)} s(n)=
\cr &=
{2^{-1}\pi} \frac{\left(l-r+\frac 12\right)}
{\left[\Gamma\left(\frac 12 -r\right)\right]^2}
\frac{\Gamma(l+1)}{\Gamma(l-2r +1) }
\frac{\Gamma(n-2r+1)}{\Gamma(n+1)} s(n) =\cr
&=
\frac 1 2  \frac{\pi \alpha_\lambda\left(\frac{l+m-n}{2}\right)
\alpha_\lambda\left(\frac{l-m+n}{2}\right)
\alpha_\lambda\left(\frac{l+m+n}{2}+\lambda\right)
\alpha_\lambda\left(\frac{l-m-n}{2}-\lambda\right)}
{ \alpha_\lambda(l) \alpha_\lambda\left(l+\lambda \right)}\ .
\label{q.10}\end{align}
This expression is symmetric in $m$ and $n$, so that
$c_{\half-r}(m,n|l) = 2R(m,\ n|l)$.
Thus, for $m$, $n$, $\lambda = \half -r$ all satisfying the conditions
stated above, and $l = m+n+2\lambda+2k$, $k$ a non-negative integer,
the statements of Theorem \ref{main} hold, and therefore this theorem
holds generally by algebraic continuation as announced in Sect. \ref{alg}.

\newpage
\section{Expansion theorems}
\label{expthms}
Theorem \ref{main} has been proved in Sect. \ref{ana}, and
it has been shown there that the conjectured identity (\ref{p.1})
holds for the values of the parameters used in that section.

Returning to the conjectured identity (\ref{p.1}), we
consider the case when $m$ and $n$ are non-negative integers
and $2\lambda > 0$ is an integer.
The function
$f(z) = (z^2-1)^{\lambda-\half}D_m^{\lambda}(z)D_n^{\lambda}(z)$ satisfies
the hypotheses of Theorem \ref{gendexp} : note that
$(z^2-1)^{\lambda-\half} = z^{2\lambda-1}(1-z^{-2})^{\lambda-\half}$ and that,
at infinity, $f(z) \sim {\rm const.\ } z^ {-(m+n+2\lambda+1)}$.
Therefore (\ref{p.1}) holds
and is uniformly convergent in any compact subset of
$\bC \setminus [-1,\ 1]$.
This uniform convergence allows the identification of the
Laurent expansions of both sides of (\ref{p.1}). Therefore
$c_{\lambda}(m,n|l)$ can again be
identified with the solution of the algebraic problem, and is
given by Theorem \ref{main}. These conclusions can be assembled
in the following theorem.

\begin{theorem}
\label{concthm}
Let $m$ and $n$ be non-negative integers, and suppose that
one of the two following conditions is satisfied:

\noindent (i) $\lambda = \half -r$, $r \ge 1$ is an integer
such that $m \ge 2r$ and $n \ge 2r$;

\noindent (ii) $2\lambda$ is a strictly positive integer;

Then
\beq
D^{\lambda}_{m}(z) D^{\lambda}_{n}(z) =
\sum_{\begin{array}{c}
\scriptstyle l = m+n+2\lambda + 2k\cr
\scriptstyle k \in \bZ,\ \ 0\le k
\end{array}}
c_\lambda(m,n|l) D^{\lambda}_{l}(z)
\label{x.1}\endq
holds with uniform convergence if $z$ remains in any
compact subset of $\bC \setminus [-1,\ 1]$, with
\begin{align}
c_\lambda(m,n|l) &=
{2^{2\lambda-1}(l+\lambda)
\Gamma(l+1)\Gamma(\lambda)^2\over i\pi^2 \Gamma(l+2\lambda)}
\int_\CC (t^2-1)^{\lambda -\half}  D^{\lambda}_{m}(t)\, D^{\lambda}_{n}(t)\,
C_l^{\lambda}(t)\,dt
\label{x.2}\\
&=\frac{\pi {\alpha_\lambda\left(\frac{l+m-n}{2}\right)}
 {\alpha_\lambda\left(\frac{l-m+n}{2}\right)}
 {\alpha_\lambda\left(\frac{l+m+n}{2}+\lambda\right)}
{\alpha_\lambda\left(\frac{l-m-n}{2}-\lambda\right)}}{ \alpha_\lambda(l)
\alpha_\lambda\left(l+\lambda \right)} \ .
\label{x.3}\end{align}
Here $\CC$  may be taken as the circle $\{z \in \bC\ :\ |z|=R\}$,
$R >1$, traversed in the positive direction.
\end{theorem}

This theorem requires $m$ and $n$ to be integers. On the other hand
Eqs (\ref{a.3}) and (\ref{y.0}) hold as identities between
formal power series in $u$ and $v$, respectively (with $f_k(x,\ y,\ \eta)$
given by (\ref{a.9})) without such restrictions. Considering again
the formal identity (\ref{y.0})
\begin{align}
&F(x,\ 1-\eta\ ;\ x+\eta\ ;\ v)
F(y,\ 1-\eta\ ;\ y+\eta\ ;\ v) =\cr
&= \sum_{k=0}^{\infty} \,f_k(x,\ y,\ \eta) v^k
F(x+y+2k,\ 1-\eta\ ;\ x+y+\eta+2k\ ;\ v)
\label{x.10}\end{align}
(with $f_k(x,\ y,\ \eta)$ given by (\ref{a.9})),
we note that the common formal expansion in powers of $v$ on the
lhs and the rhs is in fact convergent for $|v|<1$, since
the lhs is holomorphic there. This does not imply that the series
on the rhs converges. However if $a\ge 0$, $b\ge 0$ and $c >0$,
all the coefficients of $F(a,\ b\ ;\ c\ ;\ v)$ as a power series
in $v$ are positive, so that for $|v|<1$,
$|F(a,\ b\ ;\ c\ ;\ v)| \le F(a,\ b\ ;\ c\ ;\ |v|)$ and, for
$0\le v <1$, $F(a,\ b\ ;\ c\ ;\ v)$ is the least upper bound
of its partial sums. Let $x$, $y$ and $\eta$ be chosen such that
\beq
x > 0,\ \ y> 0,\ \ x+y+2\eta-1 \ge 0,\ \ 1-\eta \ge 0,\ \
x+\eta >0,\ \  y +\eta >0,\ \  x+y+\eta > 0\ .
\label{x.11}\endq
Then all the coefficients of the hypergeometric functions
appearing in (\ref{x.10}) as well as $f_k(x,\ y,\ \eta)$ are positive.
We temporarily denote $G(v)$ the lhs of (\ref{x.10}) and, for any
integers $p \ge 0$ and $q\ge 0$, $S_p(v)$ the partial sum of the
series on the rhs obtained by stopping at $k=p$, $G_q(v)$ the partial
expansion of $G(v)$ in powers of $v$ up to the power $q$, $S_{p,q}(v)$
the expansion of $S_p(v)$ in powers of $v$ up to the power $q$.
For $0\le v <1$, $G(v) = \sup_q G_q(v)$,
$S_p(v) = \sup_q S_{p,q}(v) \le \sup_q G_q(v) = G(v)$.
Thus $S_p(v)$ is bounded so that the series on the rhs converges.
For $q \le p$, $S_{p,q}(v)$ is equal
to $G_q(v)$, so that $G_q(v) \le S_p(v) \le G(v)$, hence
the sum of the series on the rhs is equal to $G(v)$.
For $|v|<1$ and integer $p<p'$,
$|S_{p'}(v) - S_p(v)| \le S_{p'}(|v|) - S_p(|v|)$, hence the sequence
$S_p(v)$ converges to a limit which is holomorphic in the unit disk
and coincides with $G$ if $v = |v|$, hence is equal to $G$.
Note also that $|S_p(v)| \le G(|v|)$ for all $v$ in the unit disk.
The map $z \mapsto v$ given by (\ref{p.5}) maps  $E_+(L)\cup\{\infty\}$
($L \ge 1$, see (\ref{s.14})), onto the disk $\{v\ :\ |v|<L^{-2}\}$.
In terms of the variables
$m$, $n$ and $\lambda$, the conditions (\ref{x.11}) follow from
\beq
\lambda >0,\ \  m+2\lambda >0,\ \  n+2\lambda >0,\ \  m+\lambda+1 >0,\ \
n+\lambda+1 >0,\ \  m+n+2\lambda+1 \ge 0\ .
\label{x.12}\endq
We thus obtain the following theorem:

\begin{theorem}
Under the conditions (\ref{x.12}),
\beq
D^{\lambda}_{m}(z) D^{\lambda}_{n}(z) =
\sum_{\begin{array}{c}
\scriptstyle l = m+n+2\lambda + 2k\cr
\scriptstyle k \in \bZ,\ \ 0\le k
\end{array}}
c_\lambda(m,n|l) D^{\lambda}_{l}(z)
\label{x.13}\endq
holds as a convergent series for $z \in \bC \setminus (-\infty,\ 1]$,
with $c_\lambda(m,n|l)$ given by (\ref{x.3}).
\end{theorem}
We emphasize that none of the parameters $m$, $n$ and $2\lambda$
has to be an integer in this theorem, but the conditions (\ref{x.12})
must be satisfied.
The proof of this theorem can be slightly
expanded to show that the convergence of (\ref{x.13}) actually holds
in the sense of functions with tempered behavior in
$\bC \setminus (-\infty,\ 1]$,
so that the conclusion also holds for the boundary values of both
sides in (\ref{x.13}).

\subsection{K\"all\'en-Lehmann weights}

Returning to the $d$-dimensional Anti-de-Sitter space-time $X_d$ (or its
covering $\wt X_d$), with $d \ge 2$, setting $\lambda = (d-1)/2$,
and taking into account the formulae
(\ref{s.6}, \ref{s.7}, and \ref{s.10}), we obtain the following
results:

\begin{theorem}
\label{k-l}
Let $m$ and $n$ be integers satisfying the conditions (\ref{x.12}).
\beq
W_{m+{d-1\over 2}}(z_1,\ z_2)\,W_{n+{d-1\over 2}}(z_1,\ z_2) =
\sum_{\begin{array}{c}
\scriptstyle l = m+n+d-1 + 2k\cr
\scriptstyle k \in \bZ,\ \ 0\le k
\end{array}} \rho(l;\ m,\ n)\,W_{l+{d-1\over 2}}(z_1,\ z_2)\ ,
\label{x.15}\endq
with
\begin{align}
\rho(l;\ m,\ n) &= {\Gamma(\lambda)\over 2\pi^{\lambda}}
\frac{{\alpha_\lambda\left(\frac{l+m-n}{2}\right)}
 {\alpha_\lambda\left(\frac{l-m+n}{2}\right)}
 {\alpha_\lambda\left(\frac{l+m+n}{2}+\lambda\right)}
{\alpha_\lambda\left(\frac{l-m-n}{2}-\lambda\right)}}{ \alpha_\lambda(l)
\alpha_\lambda\left(l+\lambda \right)} \ ,\cr
\lambda &={d-1\over 2}\ .
\label{x.16}\end{align}
Here $z_1 \in \TT_{1-}$, $z_1 \in \TT_{1+}$, and the convergence holds
in the sense of holomorphic functions with tempered behavior in
$\TT_{1-} \times \TT_{1+}$, so that the above equation extends to
the boundary values $\WW$ of the functions $W$.
The same equation holds in the case of $\wt X_d$, with
$z_1 \in \wt \TT_{1-}$ and $z_2 \in \wt \TT_{1+}$, and with $m$ and $n$
not necessarily integers, but satisfying the conditions (\ref{x.12}).
\end{theorem}
While $(l,\ m,\ n) \mapsto \rho(l;\ m,\ n)$ will always denote
the meromorphic function defined by (\ref{x.16}), the sum in
(\ref{x.15}) begins at $l= m+n+d-1$.
This spectral property is in sharp contrast to the
situation in the de Sitter case. It reflects the fact that a genuine
positive-energy condition has been imposed in the AdS case.

\newpage
\section{Some applications}
\label{appli}
In this section the radius $R$ of $X_d$ will no longer be fixed as 1,
and the AdS quadric with radius $R$ given by (\ref{s.1})
will be denoted $X_d(R)$.
In this case, for the free Klein-Gordon field $\phi$ labelled by
$n+(d-1)/2$,
\begin{align}
&(\Omega,\ \phi(x_1)\phi(x_2)\,\Omega) =
\WW_{n+{d-1\over 2}}(x_1,\ x_2) =
\lim_{\begin{array}{c}
\scriptstyle z_1 \in \TT_{1-},\ \ z_2 \in \TT_{1+}\\
\scriptstyle z_1\rightarrow x_1,\ \ z_2\rightarrow x_2 \end{array}}
W_{n+{d-1\over 2}}(z_1,\ z_2)\ ,
\label{z.20}\\
&W_{n+{d-1\over 2}}(z_1,\ z_2) = R^{2-d}
w_{n+{d-1\over 2}}\left ({z_1\cdot z_2\over R^2} \right)\ ,
\label{z.21}\end{align}
where $w_{n+(d-1)/2}$ is given by (\ref{s.10}). We keep the formula
(\ref{x.16}) so that the rsh of (\ref{x.15}) now acquires a factor
$R^{2-d}$.

In a Minkowski, de Sitter or Anti-de-Sitter space, we
consider three commuting Klein-Gordon fields $\phi_0$, $\phi_1$
and $\phi_2$ operating in the same Fock space $\FF$
(with vacuum $\Omega$),
and denote $\LL(x) = \phi_0(x)\phi_1(x)\phi_2(x)$.
The fields have masses $m_j$ or, in the AdS case,
parameters $n_j+(d-1)/2$, $j=0,\ 1,\ 2$.
Let $f_0$ be a test-function and
$\psi_0 = \int f_0(x) \phi_0(x)\Omega\,dx$. Let $E_{1,2}$
be the projector on the subspace spanned by the states of the form
$\int \vhi(x_1,\ x_2) \phi_1(x_1)\phi_2(x_2)\,\Omega\,dx_1\,dx_2$.
If an interaction of the form $I_g = \int \coupl g(x)\,\LL(x)\,dx$
is introduced, with a coupling constant $\coupl$, and
with $g$ a real, rapidly decreasing, smooth switching-off factor,
the lowest order transition probability from $\psi_0$ to any
state in $E_{1,2}\FF$ is given by
\begin{align}
&{(\psi_0,\ I_g E_{1,2}I_g \psi_0)\over (\psi_0,\ \psi_0)} = \cr
&= {\coupl^2\int \ovl{f_0(x)} g(u) g(v) f_0(y)\,
\WW_{m_0}(x,\ u)\,\WW_{m_1}(u,\ v)\,\WW_{m_2}(u,\ v)\,
\WW_{m_0}(v,\ y)\,dx\,du\,dv\,dy\over
\int \ovl{f_0(x)}\,\WW_{m_0}(x,\ y)\,f_0(y)\,dx\,dy}\ .
\label{z.1}\end{align}
Attempting to take the ``adiabatic limit'' of this expression,
i.e. its limit as $g$ tends to 1,  leads, in Minkowski or
de Sitter space-time, to a divergence for which the traditional remedy is
the Fermi golden rule. This requires involved computations in
the de Sitter case \cite{bros2,bros3}.
It will be seen below that the corresponding
calculation is considerably easier in the case of the
AdS space-time. The question of its physical interpretation
is, however, considerably more difficult. It seems nevertheless
worth giving it here as a simple application of Theorem \ref{k-l}.
Another ingredient is the ``projector
identity'' for $X_d(R)$ (analogous to a similar property in the Minkowskian
and de Sitter case \cite{bros2}), given by the following theorem.

\begin{theorem}[Projector identity]
\label{projid}
Let $n_1$, $n_2$ and $d$ be integers satisfying $d \ge 2$,
$n_1+d-1 >0$, $n_2+d-1>0$ and $n_1+n_2+d-1>0$. Then
\beq
\int_{X_d} W_{n_1+{d-1\over 2}}(z_1,\ u)\,
W_{n_2+{d-1\over 2}}(u,\ z_2)\,du =
{2\pi R^2 \over (2n_1+d-1)}\delta_{n_1n_2} W_{n_1+{d-1\over 2}}(z_1,\ z_2)\ ,
\label{z.2}\endq
Here $z_1 \in \TT_-$ and $z_2 \in \TT_+$, and $du$ denotes the
standard invariant measure on $X_d$,
$du = 2R\delta(u\cdot u -R^2)du^0\ldots du^d$. The convergence is absolute
and uniform when $(z_1,\ z_2)$ remains in a compact subset of
$\TT_-\times \TT_+$, and, in fact the convergence takes place in the
space of functions holomorphic with tempered behavior in
$\TT_-\times \TT_+$,
and the equation continues to hold for the boundary values
$\WW_{n_j+{d-1\over 2}}$ of the functions $W_{n_j+{d-1\over 2}}$.
\end{theorem}
This theorem is proved in Appendix \ref{adsproj}.
Note that it gives another proof of the positive-definiteness
of $W_{n+{d-1\over 2}}(z_1,\ z_2)$ for integer $n$ satisfying
$2n+d-1>0$.

Let $n_0$, $n_1$, and $n_2$ be integers such that
\begin{align}
&n_1+d-1 >0,\ \ n_2+d-1 >0,\ \ n_1 +{d-1\over 2} +1 > 0,\ \
n_2 +{d-1\over 2} +1 > 0,
\label{z.3}\\
&n_0+d-1 >0,\ \ n_0+n_1+n_2+2(d-1)>0\ .
\label{z.3.1}\end{align}
((\ref{z.3}) implies $n_1+n_2+d+1>0$ hence $n_1+n_2+d\ge 0$.)
Let $z_1 \in \TT_{1-}$, $z_2 \in \TT_{1+}$, $g_1$ and $g_2$ be two
smooth functions with rapid decrease on $X_d$.
By Theorem \ref{k-l},
\begin{align}
& \int_{X_d\times X_d} W_{n_0+{d-1\over 2}}(z_1,\ u)\,W_{n_1+{d-1\over 2}}(u,\ v)\,
W_{n_2+{d-1\over 2}}(u,\ v)\,W_{n_0+{d-1\over 2}}(v,\ z_2)\,
g_1(u)\,du\,g_2(v)\,dv = \cr
&= \sum_{\begin{array}{c}
\scriptstyle l = n_1+n_2+d-1 + 2k\cr
\scriptstyle k \in \bZ,\ \ 0\le k
\end{array}} R^{2-d}\rho(l;\ n_1,\ n_2) \times \cr
& \times
\int_{X_d\times X_d} W_{n_0+{d-1\over 2}}(z_1,\ u)\,\WW_{l+{d-1\over 2}}(u,\ v)\,
W_{n_0+{d-1\over 2}}(v,\ z_2)\,g_1(u)\,du\,g_2(v)\,dv \ .
\label{z.4}\end{align}
Note that by arguments similar to the proof of Theorem \ref{projid} the
integral in the lhs of (\ref{z.4}) is absolutely convergent even
if $g_1$ and $g_2$ are set equal to 1.
In the last integral, we can let $g_1$ tend to 1 and execute the integration
over $u$ by applying the projector identity, then let $g_2$ tend to 1
and similarly execute the integration over $v$. Thus
\begin{align}
 \int_{X_d\times X_d} W_{n_0+{d-1\over 2}}(z_1,\ u)\,&W_{n_1+{d-1\over 2}}(u,\ v)\,
W_{n_2+{d-1\over 2}}(u,\ v)\,W_{n_0+{d-1\over 2}}(v,\ z_2)\,du\,dv = \cr
&= R^{6-d}\left ( {2\pi \over 2n_0+d-1} \right )^2\,\rho(n_0;\ n_1,\ n_2)\,
W_{n_0+{d-1\over 2}}(z_1,\ z_2)\ ,
\label{z.5}\end{align}
provided $n_0 - n_1-n_2-d+1$ is an even non-negative integer,
the lhs being otherwise equal to zero.
Under the same condition
the limit as $g$ tends to 1 of (\ref{z.1}) is given by
\beq
{\rm Prob.\ }(\psi_0 \rightarrow n_1,\ n_2) \bydef
\lim_{g \rightarrow 1}
{(\psi_0,\ I_g E_{1,2}I_g \psi_0)\over (\psi_0,\ \psi_0)} =
R^{6-d}\left ( {2\pi \coupl \over 2n_0+d-1} \right )^2\,
\rho(n_0;\ n_1,\ n_2)\ .
\label{z.6}\endq
As in the de Sitter case, this expression is independent of the
initial wave-function $f_0$. There has been
no necessity for using the Fermi golden rule. In order do so nevertheless,
i.e. take the ``time-average'' of this ``transition probability'',
we need to divide the expression in (\ref{z.6}) by some plausible
``total time'' of the form $K_0 R$. The result is
\beq
\hbox{``Time-average''\ } ({\rm Prob.\ }(\psi_0 \rightarrow n_1,\ n_2)) =
{4\pi^2 R^{5-d}\coupl^2 \over K_0\,(2n_0+d-1)^2} \rho(n_0;\ n_1,\ n_2)\ .
\label{z.7}\endq

\subsection{Minkowskian limits}
\label{minklim}

In this subsection,
$n$ will not necessarily be an integer and $w_{n+(d-1)/2}$ is regarded
as holomorphic in $\bC \setminus (-\infty,\ 1]$.

If the origin of coordinates in $\bR^{d+1}$ is transported
to the point $Re_d = (0,\ \ldots,\ R) \in X_d(R)$, and the radius $R$
is allowed to tend to $+\infty$, the translated quadric
$X_d(R)-Re_d$ tends to the Minkowski subspace $M_d = \{x\ :\ x^d =0\}$.
The Klein-Gordon field on $X_d(R)-Re_d$ with parameter $n = mR > 0$
can be considered to tend to the Klein-Gordon field on $M_d$. Let
indeed
\beq
z_1 = Re_d,\ \ \ z_2 = R\sin(t/R)e_0 + R\cos(t/R)e_d\ ,\ \ \
\Im t >0\ .
\label{z.22}\endq
It can be shown that
\begin{align}
\lim_{R\rightarrow +\infty} R^{2-d}
w_{mR+{d-1\over 2}}\left( {z_1\cdot z_2 \over R^2} \right ) &=
{i^{d-1}\over 4(2\pi)^{{d\over 2}-1}} m^{d-2} (mt)^{1-{d\over 2}}
H_{{d\over 2}-1}^{(1)}(mt)\cr
&= \wh w_{{\rm \ Minkowski},\ m}(t^2)\ .
\label{z.23}\end{align}
Here $\wh w_{{\rm \ Minkowski},\ m}$ is holomorphic in $\bC\setminus \bR_+$
and the free Klein-Gordon field
$\phi_{{\rm \ Minkowski},\ m}$ of mass $m$ on $M_d$ satisfies
\beq
(\Omega, \phi_{{\rm \ Minkowski},\ m}(r_1)\,
\phi_{{\rm \ Minkowski},\ m}(r_2)\,\Omega) =
\lim_{\begin{array}{c}
\scriptstyle \Im u_1 \in V_{-},\ \ \Im u_2 \in V_{+}\\
\scriptstyle u_1\rightarrow r_1,\ \ u_2\rightarrow r_2 \end{array}}
\wh w_{{\rm \ Minkowski},\ m}((u_1-u_2)^2)\ .
\label{z.24}\endq
It is therefore interesting to consider the behavior of the
K\"all\'en-Lehmann weight (\ref{x.16}) and the expression (\ref{z.6})
in the same limit, as it was done in \cite{bros2,bros3} in the de Sitter case.
By Stirling's formula, as $\Re t \rightarrow +\infty$ at fixed
$x$, $y$, and $\lambda$,
\beq
{\Gamma(t+x)\over \Gamma(t+y)} \sim t^{x-y}\ ,\ \ \ \
\alpha_\lambda(t) \sim {t^{\lambda-1}\over \Gamma(\lambda)}\ .
\label{z.25}\endq
For fixed $m_0 > 0$, $m_1 > 0$, and $m_2 > 0$, $\lambda = (d-1)/2$,
we have therefore
\begin{align}
\lim_{R \rightarrow +\infty}R^{3-d}&\rho(Rm_0\,;\ Rm_1,\ Rm_2) =
\theta(m_0-m_1-m_2)
{1\over 2^{2d-5}\pi^{d-1\over 2}m_0^{d-3}\Gamma \left ({d-1\over 2}\right )}
\times \cr
&\times
\left [(m_0-m_1+m_2)(m_0+m_1-m_2)(m_0+m_1+m_2)(m_0-m_1-m_2)
\right ]^{d-3\over 2}\cr
&= 4m_0\,\rho_{\rm Min}(m_0^2\,;\ m_1,\ m_2)\ .
\label{z.26}\end{align}
If we set similarly $n_j = Rm_j$ in (\ref{z.7}), this expression tends,
as $R \rightarrow +\infty$, to
\beq
{4\pi^2 \coupl^2 \over K_0\,m_0}\rho_{\rm Min}(m_0^2\,;\ m_1,\ m_2)\ .
\label{z.27}\endq
With the choice $K_0= 4\pi$, (\ref{z.27}) becomes equal to an analogous
quantity in Minkowski QFT, i.e. the inverse lifetime of a particle
of mass $m_0$ decaying into two particles of masses $m_1$ and $m_2$
in its rest-frame (see \cite{bros2,bros3}).

\newpage
\appendix

\section{Appendix. Jacobi and ultraspherical functions}
\label{jac}
\subsection{Jacobi polynomials and functions of the second kind}

The major part of this and the next subsections is taken
from \cite{S}. In both subsections, $n \ge 0$ is an integer.
For arbitrary complex $\alpha$ and $\beta$, the Jacobi
polynomial $P_n^{(\alpha,\beta)}$ is given by
\begin{align}
P_n^{(\alpha,\beta)}(x) &= \left ({n+\alpha \atop n} \right )
F\left (-n,\ n+\alpha+\beta+1\ ;\ \alpha+1\ ;{1-x\over 2} \right ) =\cr
&= {1\over n!} \sum_{p=0}^n \left ({n \atop p} \right )
(n+\alpha+\beta+1)\ldots (n+\alpha+\beta+p)\times\cr
&\times (\alpha+p+1)\ldots (\alpha+n)
\left ( {x-1 \over 2} \right )^p\ .
\label{j.1}\end{align}
(\cite[(4.21.2), p. 62]{S}). By definition,
\beq
\bin{a}{b} = {\Gamma(a+1)\over \Gamma(b+1)\Gamma(a-b+1)}
\label{j.2}\endq
Thus $P_n^{(\alpha,\beta)}(x)$ is a polynomial in $x$, $\alpha$ and
$\beta$.
Rodrigues' formula (\cite[(4.3.1), p. 67]{S}),
\beq
(1-x)^\alpha (1+x)^\beta P_n^{(\alpha,\beta)}(x) =
{(-1)^n\over 2^n n!} \left ({d\over dx} \right )^n
(1-x)^{n+\alpha} (1+x)^{n+\beta}\ ,
\label{j.3}\endq
may be taken as another definition.
For $\Re (\alpha+n) > -1$ and $\Re(\beta+n) > -1$,  and excluding the case
$n=0$ and $\alpha+\beta+1=0$, the Jacobi function of the second kind
$Q_n^{(\alpha,\beta)}(z)$ is given by
\begin{align}
(z-1)^{\alpha}(z+1)^{\beta}Q_n^{(\alpha,\beta)}(z)
&= \cr
&2^{-n-1}\int_{-1}^1 (1-t)^{n+\alpha}(1+t)^{n+\beta}(z-t)^{-n-1}\,dt\ .
\label{j.7}\end{align}
$Q_n^{(\alpha,\beta)}$ extends as a function holomorphic in
$\bC\setminus (-\infty,\ 1]$, and
$Q_n^{(\alpha,\beta)}(z) \sim z^{-n-\alpha-\beta-1}$ as $z \rightarrow \infty$.
(\cite[(4.61.1), p.~73]{S}).  Another representation is (\cite[p.~74]{S}) :
\begin{align}
Q_n^{(\alpha,\beta)}(z) &= 2^{n+\alpha+\beta}
{\Gamma(n+\alpha+1)\Gamma(n+\beta+1)\over \Gamma(2n+\alpha+\beta+2)}
(z-1)^{-n-\alpha-1}(z+1)^{-\beta}\times\cr
&\times F\left ( n+\alpha+1,\ n+1\ ;\ 2n+\alpha+\beta+2\ ;
{2\over 1-z} \right )\ .
\label{j.9}\end{align}
Eq. (\ref{j.9}) is derived from Eq (\ref{j.7}) and they both
provide the same analytic extension to complex values of $n$,
$\alpha$ and $\beta$.

\subsection{Ultraspherical functions}
\label{ultra}

For a positive integer $n$ and a complex $\lambda$, the
Gegenbauer polynomial $C_n^\lambda \equiv P_n^{(\lambda)}$ is defined as
\beq
C_n^\lambda(x) = U(n,\ \lambda) P_n^{(\lambda-\half,\ \lambda-\half)}(x)\ ,
\label{u.1}\endq
with
\beq
U(n,\ \lambda) =
{\Gamma(\lambda+\half)\Gamma(n+2\lambda)\over
\Gamma(2\lambda)\Gamma(n+\lambda+\half)} =
{2^{1-2\lambda} \pi^{1/2} \Gamma(n+2\lambda)\over
\Gamma(\lambda)\Gamma(n+\lambda+\half)}\ .
\label{u.4.1}\endq
Hence
\begin{align}
C_n^\lambda(x) &= {\Gamma(n+2\lambda)\over
\Gamma(n+1)\Gamma(2\lambda)} F \left ( -n,\ n+2\lambda\ ;\
\lambda+\half\ ;\ {1-x\over 2}\right )
\label{u.2}\\
&= {2^{\half-\lambda}\Gamma(\half)\Gamma(n+2\lambda)
(x^2-1)^{{1\over 4}-{\lambda\over 2}}\over
\Gamma(n+1)\Gamma(\lambda)}
P_{n-\half+\lambda}^{\half-\lambda}(x)\ .
\label{u.3}\end{align}
Here $P_\nu^\mu$ is the Legendre function (\cite[3.2 (3) p. 122]{HTF1}).
Rodrigues's formula gives
\beq
C_n^\lambda(x) = {2^{1-2\lambda}\pi^{1/2} \Gamma(n+2\lambda)(-1)^n\over
\Gamma(\lambda)\Gamma(n+\half+\lambda)2^n n!}
(1-x^2)^{\half-\lambda} \left ({d\over dx} \right )^n (1-x^2)^{n+\lambda-\half}\ .
\label{u.4}\endq
The Gegenbauer polynomials have a generating function
(\cite[3.15.1(1), p. 175]{HTF1}, \cite[p.~83]{S}) :
\beq
\sum_{n=0}^\infty C_n^\lambda(x)\,h^n = (1-2hx+h^2)^{-\lambda}\ .
\label{u.4.2}\endq
This shows that $\lambda \mapsto C_n^\lambda(x)$ is entire.

Define
\beq
D_n^\lambda(z) = U(n,\ \lambda) Q_n^{(\lambda-\half,\ \lambda-\half)}(z)\ .
\label{u.5}\endq
$(z^2-1)^{\lambda-\half} D_n^\lambda(z)$ extends to a function
holomorphic in $\bC\setminus [-1,\ 1]$. As a special case of (\ref{j.7}),
for $z$ in this cut-plane, and supposing $\Re(n+2\lambda)>0$
(which implies $\Re(n+\lambda-\half) > -1$ for $n \ge 0$),
\begin{align}
(z^2-1)^{\lambda-\half}D_n^\lambda(z) &=
{(-2)^{-n-1}2^{1-2\lambda}\pi^{1/2}\Gamma(n+2\lambda)\over
\Gamma(\lambda)\Gamma(n+\lambda+\half)}\,
\int_{-1}^1 (1-t^2)^{n+\lambda-\half}(t-z)^{-n-1}\,dt
\label{u.5.1}\\
&= {(-2)^{-n-1}2^{1-2\lambda}\pi^{1/2}\Gamma(n+2\lambda)\over
n!\Gamma(\lambda)\Gamma(n+\lambda+\half)}\,
\left ( {d\over dz} \right )^n
\int_{-1}^1 (1-t^2)^{n+\lambda-\half}(t-z)^{-1}\,dt\ .
\label{u.5.2}\end{align}
Therefore, in the sense of tempered distributions, on the real axis,
\begin{align}
& {\rm disc\,} \left [ (x^2-1)^{\lambda-\half}D_n^\lambda(x) \right ] =
{2i\pi(-2)^{-n-1}2^{1-2\lambda}\pi^{1/2}\Gamma(n+2\lambda)\over
n!\Gamma(\lambda)\Gamma(n+\lambda+\half)}\times\cr
& \times \left ( {d\over dx} \right )^n \left [ \theta(x+1)\,\theta(1-x)\,
(1-x^2)^{n+\lambda-\half} \right ]\ .
\label{u.5.3}\end{align}
(\ref{j.9}) gives, together with \cite[3.2 (37) p.~132]{HTF1},
and \cite[3.3.1 (2) p.~140]{HTF1},
\begin{align}
D_n^\lambda(x) &= {2^{\half -\lambda} \pi^{1/2} \Gamma(n+2\lambda)
(x^2-1)^{{1\over 4}-{\lambda\over 2}}\over
\Gamma(\lambda)\Gamma(n+1)}\, e^{i\pi(\lambda-\half)}
Q_{n+\lambda-\half}^{\half -\lambda}(x)
\label{u.8}\\
&= {2^{\half -\lambda} \pi^{1/2} (x^2-1)^{{1\over 4}-{\lambda\over2}}\over
\Gamma(\lambda)}\, e^{i\pi(\half-\lambda)}
Q_{n+\lambda-\half}^{\lambda-\half}(x)\ .
\label{u.8.1}\end{align}
Eq. (\ref{u.8}) and \cite[3.2 (8) p.~122]{HTF1}, or Eq. (\ref{u.8.1}) and
\cite[3.2 (5) p.~122]{HTF1},  give
\beq
D_n^\lambda(x) = {\pi\Gamma(n+2\lambda)\over
\Gamma(\lambda)\Gamma(n+\lambda+1)} (2x)^{-n -2\lambda}
F \left ( {n+2\lambda\over 2},\ {n+2\lambda+1\over 2}\ ;\
 n+\lambda+1\ ;\ {1\over x^2} \right )\ .
\label{u.9}\endq
This formula, (\ref{u.5.1}), (\ref{u.8}), (\ref{u.8.1}), and (\ref{s.12})
all provide the same analytic extension of $D_n^\lambda$ to complex values
of $n$.

The Legendre function of the second kind ``on the cut'', i.e. on
$(-1,\ 1)$, are defined as follows:
\beq
\Q^{\mu}_{\nu}(x) =  \frac 12 e^{-i\pi\mu}\left[e^{-\frac{i\pi\mu}2}
Q^{\mu}_{\nu}(x+i0)
+ e^{\frac{i\pi\mu}2} Q^{\mu}_{\nu}(x-i0)\right].
\label{u.9.1}\endq
It follows that
\begin{align}
 {\D}^{\lambda}_n(x)&= \frac 12 \left[{D}^{\lambda}_n(x+i0)+
{D}^{\lambda}_n(x-i0)\right]  =\cr
&=    \frac{\pi^{1/2}}{2^{ \lambda + \frac 12}\Gamma(\lambda)}
(1-x^2)^{{\frac 14 -\frac\lambda 2}} e^{-i\pi(\lambda-\frac 12)}
\left[ e^{-i\pi(\frac\lambda 2-\frac 14)}
Q^{\lambda-\frac 12}_{n +\lambda - \frac 12}(x+i0)
+e^{i\pi(\frac\lambda 2-\frac 14)}
Q^{\lambda-\frac 12}_{n +\lambda - \frac 12}(x-i0)\right] \cr
&=
\frac{\pi^{1/2}}{2^{ \lambda - \frac 12}\Gamma(\lambda)}
(1-x^2)^{{\frac 14 -\frac\lambda 2}}
\Q^{\lambda - \frac 12}_{n +\lambda - \frac 12}(x)
\label{u.9.2}\end{align}
Also note the following formulae from \cite[p. 143]{HTF1}, valid
for $-1 < x < 1$ :
\begin{align}
&\P_\nu^\mu(x) = e^{i\pi\mu\over 2} P_\nu^\mu(x+i0) =
e^{-i\pi\mu\over 2} P_\nu^\mu(x-i0) =\cr
&= {1\over \Gamma(1-\mu)} \left ( {1+x \over 1-x} \right )^{\mu \over 2}
F \left ( -\nu,\ \nu+1\ ;\ 1-\mu\ ;\ {1-x \over 2} \right )\ .
\label{u.9.3}\end{align}
\beq
\Gamma(\nu+\mu+1)\Q_\nu^{-\mu}(x) = \Gamma(\nu-\mu+1) \left [
\Q_\nu^{\mu}(x)\cos(\mu\pi) + {\pi\over 2} \P_\nu^{\mu}(x)\sin(\mu\pi)
\right ]\ .
\label{u.9.4}\endq
Thus it follows from (\ref{u.9.2}) and (\ref{u.9.4}) that
\beq
{\D}^{\lambda}_n(x) = {\pi^{1/2}\Gamma(n+2\lambda)
(1-x^2)^{{1\over 4}-{\lambda\over 2}} \over
2^{ \lambda - \half}\Gamma(n+1) \Gamma(\lambda)} \left [
\Q_{n+\lambda-\half}^{\half -\lambda}(x) \cos(\pi(\half -\lambda)) +
{\pi \over 2} \P_{n+\lambda-\half}^{\half -\lambda}(x)
\sin(\pi(\half -\lambda)) \right ]\ .
\label{u.9.5}\endq

\subsection{The special case of $\lambda = \half -r$ with integer $r\ge 0$}
\label{jacspec}

In this subsection, $r$ and $n$ are integers such that
$0 \le 2r \le n$. Eq. (\ref{u.4}) takes the form
\beq
C_n^{\half-r}(x) = {(-2)^{-n}\over n!} U(n,\ \half-r)
(1-x^2)^r \left ({d\over dx} \right )^n (1-x^2)^{n-r}\ .
\label{v.1}\endq
$U(n,\ \lambda)$ has been defined in (\ref{u.4.1}).
Since $n-r \ge 0$, this displays the fact that the polynomial
$C_n^{\half-r}(x)$ is divisible by $(1-x^2)^r$.

If $F$ is a holomorphic function of tempered behavior in the
complement of the real axis, and $\vhi$ a function holomorphic in
a complex neighborhood of the real axis, then
\beq
\vhi(x)\,{\rm disc\,} F(x) = {\rm disc\,} [\vhi(x) F(x)].
\label{v.3}\endq
Applying this to $\vhi(x) = (x^2-1)^r$ and
$F(x) = (x^2-1)^{-r}D_n^{\half-r}(x)$, we obtain from (\ref{u.5.3})
\beq
{\rm disc\,}D_n^{\half-r}(x) = {2i\pi(-2)^{-n-1}\over n!} U(n,\ \half-r)
(x^2-1)^r \left ( {d\over dx} \right )^n \left [ \theta(x+1)\,\theta(1-x)\,
(1-x^2)^{n-r} \right ]\ .
\label{v.4}\endq
This can be rewritten as
\begin{align}
{\rm disc\,}D_n^{\half-r}(x) &=
\theta(x+1)\,\theta(1-x)\,{(-i\pi)(-2)^{-n}\over n!}
U(n,\ \half-r)(x^2-1)^r \left ( {d\over dx} \right )^n (1-x^2)^{n-r} =
\label{v.5}\\
&=  (-i\pi)(-1)^r\theta(x+1)\,\theta(1-x)\, C_n^{\half-r}(x)\ .
\label{v.6}\end{align}
To see that (\ref{v.5}) follows from (\ref{v.4}), we first note
that, by Leibniz's rule,
\begin{align}
(1-x^2)^r &\left ( {d\over dx} \right )^n \left [ \theta(x+1)\,\theta(1-x)\,
(1-x^2)^{n-r} \right ] = \cr
& u^rv^r
\left ( {\partial \over \partial u} - {\partial \over \partial v}
\right )^n
\left [ \theta(u) u^{n-r} \theta(v) v^{n-r} \right ]\,
\Big |_{u = x+1,\ v = 1-x}\ .
\label{v.7}\end{align}
It is then easy to check that if $s$ is an integer such that $0\le s\le n$,
\beq
u^r(d/du)^s\theta(u)\,u^{n-r} = \theta(u)\,u^r(d/du)^s\,u^{n-r}\ .
\label{v.7.1}\endq
As a consequence, if $z\in \bC\setminus [-1,\ 1]$,
\beq
D_n^{\half-r}(z) = {(-1)^{r+1}\over 2}
\int_{-1}^1 {C_n^{\half-r}(t) \over t-z}\,dt\ .
\label{v.8}\endq
Indeed both sides have the same discontinuity and vanish at infinity.
This can be rewritten (see \cite[ p.~77]{S}) as
\beq
D_n^{\half-r}(z) = {(-1)^{r+1}\over 2}
\int_{-1}^1 {C_n^{\half-r}(t) - C_n^{\half-r}(z)\over t-z}\,dt
+ {(-1)^{r+1}C_n^{\half-r}(z)\over 2}
\int_{-1}^1 {dt \over t-z}\ .
\label{v.9}\endq
The first term is a polynomial of degree $n-1$ in $z$. The second
is equal to
\beq
{(-1)^{r+1}C_n^{\half-r}(z)\over 2} \log \left ( {z-1\over z+1} \right ).
\label{v.10}\endq
Since $C_n^{\half-r}(z)$ is divisible by $(z^2-1)^r$ this shows
that the boundary values of $D_n^{\half-r}(z)$ on the real axis
are continuous if $r \ge 1$, and in fact belong to $\CC^{r-1}$.

Formulae (\ref{u.9.2}) and (\ref{u.9.5}) become:
\begin{align}
\D_n^{\half-r}(x) &= {2^r \pi^{1/2}(1-x^2)^{r/2} \over
\Gamma(\half-r) } \Q_{n-r}^{-r}(x)
\label{v.11}\\
&= {2^r \pi^{1/2}\Gamma(n-2r+1)(1-x^2)^{r/2} \over
\Gamma(n+1)\Gamma(\half-r) } \Q_{n-r}^r(x)\,\cos(\pi r)\ .
\label{v.12}\end{align}

\subsection{Expansion of holomorphic functions in terms
of ultraspherical functions}
\label{uexp}
Expansions of holomorphic functions in series of Legendre
polynomials and functions of the second kind are classical.
Recall that the Legendre polynomials $P_n = P_n^{(0,0)}$
and Legendre functions of the second kind $Q_n = Q_n^{(0,0)}$ are given by
\beq
P_n(z) = {1\over 2^n n!} \left ( {d\over dz} \right )^n (z^2-1)^n\ ,
\ \ \ \ \
Q_n(z) = -\half \int_{-1}^1 {P_n(t) \over t-z}\,dt
\label{l.3}\endq
The following theorem is classical (the notations $E(L)$, $E_\pm(L)$
have been defined in Sect. \ref{prelim}).

\begin{theorem}
\label{lqexp}
Let $F$ be holomorphic at infinity, with $F(\infty) =0$.
Then
\beq
F(z) = \sum_{n=0}^\infty b_n Q_n(z)
\label{l.7}\endq
with
\beq
b_n = {2n+1\over 2i\pi} \int_\CC F(z)\,P_n(z)\,dz\ .
\label{l.8}\endq
The contour $\CC$ may be taken to be
any $E(L)$ such that $F$ is holomorphic in $E_+(L-\veps)$,
$L> L-\veps > 1$. The expansion (\ref{l.7}) converges uniformly
on any compact subset of the exterior of the smallest ellipse
$E(L_0)$ in the exterior of which $F$ is regular.
\end{theorem}

This theorem has a  generalization
to Jacobi polynomials and functions of the second kind: see
Theorems 9.2.1 and 9.2.2 in \cite[pp.~251-252]{S}. Applying the second of
these theorems to the special case of ultraspherical functions yields:

\begin{theorem}
\label{gendexp}
Assume $\lambda > 0$.
Let $f$ be holomorphic in a neighborhood of $\infty$, with $f(\infty) = 0$.
Then
\beq
f(y) = (y^2-1)^{\lambda-\half} \sum_{n=0}^\infty b_n D_n^{\lambda}(y)\ .
\label{u.14}\endq
This expansion is convergent in the exterior of the smallest ellipse
with foci $\pm 1$ in the exterior of which $f$ is holomorphic.
The sum of the semi-axes of this ellipse is
\beq
\limsup_{n \rightarrow \infty} |b_n|^{1/n}\ .
\label{u.15}\endq
The coefficients $b_n$ are given by
\beq
b_n = {2^{2\lambda-1}(n+\lambda)
\Gamma(n+1)\Gamma(\lambda)^2\over i\pi^2 \Gamma(n+2\lambda)}
\int f(x) C_n^{\lambda}(x)\,dx\ ,
\label{u.16}\endq
where the integral is over any larger ellipse.
\end{theorem}

As stated here, this theorem does not apply to the case
$\lambda = \half-r$, $r\in \bN$ which we will need to consider.
Although it would be possible to extend the proof of
Theorem \ref{gendexp} at the cost of some effort,
we will rely on an elementary application of
Theorem \ref{lqexp} which will suffice for our needs.

Let $F$ be a function holomorphic in $\bC\setminus [-1,\ 1]$ and at
infinity, i.e. having a convergent Laurent expansion
\beq
F(z) = \sum_{n=0}^\infty c_n z^{-n}\ ,\ \ \ \
c_n = {1\over 2i\pi} \int_\CC z^{n-1}F(z)\,dz\ .
\label{e.1}\endq
We suppose that for a certain integer $N > 1$, $c_n$
vanishes for all $n < N$, i.e. the Laurent series starts at $n=N$.
In eq. (\ref{e.1}) the contour $\CC$ may be the circle
$\{z\ :\ |z| = R\}$, with $R >1$,with the positive orientation,
or any smooth closed contour homotopic to this circle in the
cut-plane $\bC\setminus [-1,\ 1]$. We can define, for each integer
$r$ with $0\le r <N$
\beq
F^{(-r)}(z) = \sum_{n=N}^\infty {(-1)^r c_n \over
(n-1)\ldots (n-r) z^{n-r}}\ .
\label{e.3}\endq
Thus $F^{(0)} = F$ and, if $r >0$,
\beq
F^{(-r)}(z) = \int_\infty^z F^{(-r+1)}(t)\,dt\ ,
\label{e.4}\endq
the integral being over any arc in $\bC\setminus [-1,\ 1]$.
For a fixed $r$ ($1\le r <N$), Theorem \ref{lqexp} applied
to $F^{(-r)}$ gives
\beq
F^{(-r)}(z) = \sum_{n=0}^\infty b_n Q_n(z)\ ,
\label{e.5}\endq
converging uniformly on any compact subset of $\bC\setminus [-1,\ 1]$,
with
\beq
b_n = {2n+1\over 2i\pi} \int_\CC F^{(-r)}(z)\,P_n(z)\,dz\ .
\label{e.6}\endq
Here $\CC$ may be $E(L)$ or $\{z\ :\ |z| = L\}$ for any $L>1$.
Since $F^{(-r)}(z) \sim {\rm const.\ } z^{-N+r}$ at infinity,
$b_n = 0$ for $n < N-r-1$. We can now prove

\begin{theorem}
\label{dexp}
Let $N$ and $r$ be integers such that $N-r-1 \ge r \ge 1$.
Let $F$ be a function holomorphic in $\bC \setminus [-1,\ 1]$
and at infinity, with $c_n$ given by (\ref{e.1}) and
$c_n =0$ for $n<N$. Then
\beq
F(z) = \sum_{n = N-1}^\infty a_n (z^2-1)^{-r} D_n^{\half-r}(z)
\label{e.7}\endq
with
\beq
a_n = {(2n-2r+1)\Gamma(n+1) \Gamma(\half-r)^2\over
2^{2r+1}i \pi^2 \Gamma(n-2r+1)}
\int_\CC F(t)\,C_n^{\half-r}(t)\,dt\ .
\label{e.8}\endq
The convergence is uniform on any compact subset
of $\bC \setminus [-1,\ 1]$.
\end{theorem}

\noindent {\bf Proof.} Eqs. (\ref{e.5}) and (\ref{e.6}) hold, and
$b_n =0$ for $n < N-r-1$, in particular for $n< r$.
Since the series in (\ref{e.5}) is a uniformly convergent series
of holomorphic functions, it can be differentiated term by term:
\begin{align}
F(z) &= \sum_{n = N-1-r}^\infty b_n Q_n^{(r)}(z)
\label{e.9.1}\\
&= \sum_{n = N-1}^\infty b_{n-r} (z^2-1)^{-r/2} Q_{n-r}^r(z)\cr
&= \sum_{n = N-1}^\infty b_{n-r} {(-1)^r\Gamma(n+1)\Gamma(\half-r)\over
2^r\pi^{1/2}\Gamma(n-2r+1)}
(z^2-1)^{-r} D_n^{\half-r}(z)\ .
\label{e.9}\end{align}
In (\ref{e.6}), we can substitute (see \cite[3.6.1 (8) p.~149 and
3.3.1 (7) p.~140]{HTF1})
\begin{align}
P_n(z) &= \left ( {d\over dz} \right )^r (z^2-1)^{r/2} P_n^{-r}(z)
\label{e.10}\\
&= \left ( {d\over dz} \right )^r
{\Gamma(n-r+1)\over \Gamma(n+r+1)} (z^2-1)^{r/2} P_n^{r}(z)\ \ \ \
\hbox{provided \ } n \ge r\ .
\label{e.11}\end{align}
Therefore
\beq
b_{n} = {(-1)^r(2n+1)\over 2i\pi}
\int_\CC F(t)\,(t^2-1)^{r/2} P_{n}^{-r}(t)\,dt\ ,
\label{e.12}\endq
and for $n \ge 2r$,
\begin{align}
b_{n-r} &= {(-1)^r(2n-2r+1)\Gamma(n-2r+1)\over 2i\pi\Gamma(n+1) }
\int_\CC F(t)\,(t^2-1)^{r/2} P_{n-r}^{r}(t)\,dt\cr
&= {(-1)^r(2n-2r+1)\Gamma(\half-r)\over 2^{r+1}i\pi^{3/2}}
\int_\CC F(t)\,C_n^{\half-r}(t)\,dt\ .
\label{e.13}\end{align}
Substituting this into (\ref{e.9}) gives (\ref{e.8}) and proves the
theorem.

\subsection{Expansion of $\Q_{n-r}^r$ in terms of the $\P_k^r$}
\label{qrpr}
Recall that $P_\nu^0 = P_\nu$ and $Q_\nu^0 = Q_\nu$ are the
Legendre functions of the first and second kind, and that, for integer
$r \ge 1$, (see \cite[3.6.1 p.148-149]{HTF1}), for $-1<x<1$,
\beq
\P_\nu^r(x) = (-1)^r (1-x^2)^{r/2} \left ( {d\over dx} \right)^r\,
\P_\nu(x)\,\ ,
\label{v.13}\endq
\beq
\Q_\nu^r(x) = (-1)^r (1-x^2)^{r/2} \left ( {d\over dx} \right)^r\,
\Q_\nu(x)\,\ ,
\label{v.14}\endq
The Legendre polynomials $P_k = \P_k$ form an orthogonal basis of
$L^2([-1,\ 1])$ (with the Lebesgue measure) and
\beq
\int_{-1}^1 \P_k(x) \P_l(x)\,dx = {1\over k+\half} \delta_{kl}\ ,
\label{v.15}\endq
so that, for any $f,\ g \in L^2([-1,\ 1])$,
\begin{align}
&\int _{-1}^1 g(x)\,f(x)\,dx =
\sum_{k=0}^\infty f_k \int _{-1}^1 g(x)\,\P_k(x)\,dx\ ,\cr
& f_k = (k+\half) \int _{-1}^1 f(x)\,\P_k(x)\,dx
\label{v.16}\end{align}
We may also regard $f$ and $\P_k$ as distributions.
If $h$ is a $\CC^\infty$ test-function with support contained
in $(-1,\ 1)$,
\begin{align}
&\int _{-1}^1 h(x)(-1)^r(1-x^2)^{r/2}\,f^{(r)}(x)\,dx \bydef
\int _{-1}^1 \left ( \left ({d\over dx}\right )^r
\left [h(x)(1-x^2)^{r/2}\right ] \right )\,f(x)\,dx\label{v.17}\\
&= \sum_{k=0}^\infty f_k \int _{-1}^1
\left ( \left ({d\over dx}\right )^r
\left [h(x)(1-x^2)^{r/2}\right ] \right )\,\P_k(x)\,dx \label{v.18}\\
&= \sum_{k=0}^\infty f_k \int _{-1}^1 h(x)\,\P_k^r(x)\,dx\ .
\label{v.19}\end{align}
This will continue to hold if $h$ tends to a function such that
\beq
x \mapsto \left ({d\over dx}\right )^r \left [h(x)(1-x^2)^{r/2}\right ]
\label{v.20}\endq
defines an element of $L^2(\bR)$ with support in $[-1,\ 1])$.
We may in particular choose $f = \Q_N$, with $N \ge r \ge 1$.
Then $f \in L^2([-1,\ 1])$, and
\beq
f_k = (k+\half)\int _{-1}^1 \Q_N(x)\,\P_k(x)\,dx =
\left \{ \begin{array}{l}
{\displaystyle (1-(-1)^{k-N}) (k+\half)
\over \displaystyle (k-N)(k+N+1)}\ \ \ {\rm if\ } k \not= N\\
\hbox to 1cm{\hfill}\\
0\ \ \ {\rm if\ } k = N
\end{array} \right .
\label{v.21}\endq
Thus
\beq
\int_{-1}^1 h(x)\Q_N^r(x)\,dx =
\sum_{k=0}^\infty f_k \int _{-1}^1 h(x)\,\P_k^r(x)\,dx\ ,
\label{v.22}\endq
with the $f_k$ given by (\ref{v.21}). Setting $N = n-r$, with
an integer $n \ge 2r$, we obtain
\beq
\int_{-1}^1 h(x)\Q_{n-r}^r(x)\,dx =
\sum_{k=2r}^\infty a_r(k,\ n) \int _{-1}^1 h(x)\,\P_{k-r}^r(x)\,dx\ ,
\label{v.23}\endq
with
\beq
a_r(k,\ n) = a(k-r,\ n-r),\ \ \ \
a(k,\ n) = \left \{
\begin{array}{ll}
\displaystyle {1\over k-n} + {1 \over k+n+1} & {\rm if\ } k-n\ {\rm is\ odd}\\
\hbox to 1cm{\hfill}\\
0 & {\rm if\ } k-n\ {\rm is\ even}
\end{array} \right .
\label{v.24}\endq

\section{Appendix. Extension of Hs\"u's Theorem}
\label{hsu}

\begin{theorem}[Hs\"u \cite{Hs}]
\label{hsuthm}
Let $r$ be complex with $\Re r < 1$, and $n_1$, $n_2$, $n_3$ be
non-negative integers. Then the integral
\beq
\int_{-1}^{1} (1-x^2)^{-r} C_{n_1}^{\half-r}(x)\,
C_{n_2}^{\half-r}(x)\,C_{n_3}^{\half-r}(x)\,dx
\label{h.1}\endq
vanishes unless
\beq
n_j \le n_k + n_\ell,\ \ \ \ 2s \bydef n_1+n_2+n_3\ \ {\rm is\ even}
\label{h.4}\endq
for every permutation $(j,\ k,\ \ell)$ of $(1,\ 2,\ 3)$.
If the above conditions are satisfied,
\begin{align}
&\int_{-1}^{1} (1-x^2)^{-r} C_{n_1}^{\half-r}(x)\,
C_{n_2}^{\half-r}(x)\,C_{n_3}^{\half-r}(x)\,dx = \cr
&= 2^r\pi {\Gamma(s+1-2r) \over \Gamma \left (s+{3\over 2}-r \right )
\Gamma(\half -r)} \times\cr
&\times {\Gamma(s-n_1+\half -r) \over \Gamma(s-n_1+1) \Gamma(\half -r)}
{\Gamma(s-n_2+\half -r) \over \Gamma(s-n_2+1) \Gamma(\half -r)}
{\Gamma(s-n_3+\half -r) \over \Gamma(s-n_3+1) \Gamma(\half -r)}
\label{h.5}\\
& \bydef H(r\ ;\ n_1,\ n_2,\ n_3)\ .
\label{h.5.1}\end{align}

\end{theorem}

In the sequel we will take $H(r\ ;\ n_1,\ n_2,\ n_3)$ to be defined
by the meromorphic function of $r$ appearing in the rhs of (\ref{h.5})
if the conditions (\ref{h.4}) hold, and by 0 otherwise.
We abbreviate $H(r\ ;\ n_1,\ n_2,\ n_3)$ to $H(r)$
when no ambiguity arises.

\begin{remark}\rm
\label{hrem}
It is important to note that if $n_1$, $n_2$, $n_3$  are fixed
non-negative integers, then $r\mapsto H(r\ ;\ n_1,\ n_2,\ n_3)$ is holomorphic
at every integer value of $r$ such that $n_j-2r \ge 0$ for at least two
distinct values of $j=1,\ 2,\ 3$. This is obvious if the conditions
(\ref{h.4}) are not satisfied since $H(r\ ;\ n_1,\ n_2,\ n_3)= 0$
in this case.
If the conditions (\ref{h.4}) are satisfied, the three last
factors in the rhs of (\ref{h.5}) are polynomials in $r$,
while the argument of the first Gamma function is $\ge 1$.
\end{remark}

\subsection{Contour integrals}
\label{cont}

\begin{figure}[ht]
\begin{center}
\includegraphics[width=12cm]{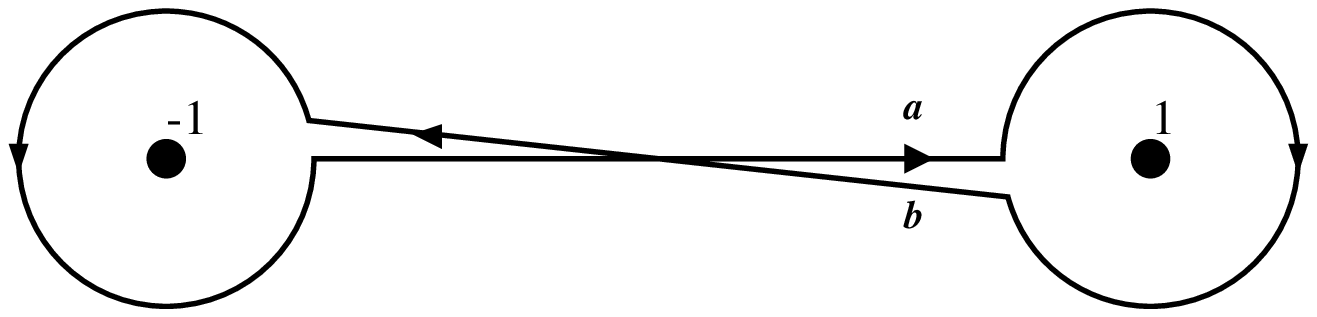}
\caption{The contour $\CC_1$} \label{fig:contmg}
\end{center}
\end{figure}

Let $\CC_1$ be the contour in Fig. \ref{fig:contmg},
which is homotopic to a figure eight.
The radii of the two circles are to be regarded as small and the
two straight segments are very close to the real axis.
Let $\Phi(x,\ r)$
be an entire function of $x$ and $r$. Let
\beq
K(r) = \int_{\CC_1} (1-x^2)^{-r}\,\Phi(x,\ r)\,dx\ .
\label{c.1}\endq
To make things more definite, we assume that the segment $a$ lies on
the real axis inside the open interval $(-1,\ 1)$, and that
on this segment $(1-x^2)^{-r} = |1-x^2|^{-r}$.
Then the contour may be considered as
a closed curve in the Riemann surface of
$z \mapsto (1-z^2)^{-r}\,\Phi(z,\ r)$. The function $K$ is entire
and can, of course, be defined with any smooth closed contour homotopic
to $\CC_1$ in that Riemann surface.

Let first $\Re r < 1$. Then the integral
\beq
I(r) = \int_{-1}^1 (1-x^2)^{-r}\,\Phi(x,\ r)\,dx
\label{c.2}\endq
exists and is holomorphic in $r$.
If the radii of the circles (centered at $1$ and $-1$) which
appear in $\CC_1$ tend to 0, their contributions to $K(r)$ tend to 0,
and the contributions of the two segments become
\begin{align}
a\ :\ & I(r)\cr
b\ :\ & -e^{-2i\pi r}I(r)\ ,
\label{c.3}\end{align}
so that, for $\Re r < 1$,
\beq
K(r) = (1-e^{-2i\pi r}) I(r) = 2ie^{-i\pi r}\sin (\pi r)\,I(r)\ .
\label{c.4}\endq
This holomorphic function of $r$ vanishes
at every integer value of $r$ in the half-plane $r < 1$.

In the case of Hs\"u's integral,
\beq
\Phi(x,\ r) =
{C}^{\frac 12 -r}_{n_1}(x) {C}^{\frac 12 -r}_{n_2}(x)
{C}^{\frac 12 -r}_{n_3}(x)\ ,
\label{c.6}\endq
where the $n_j$ are non-negative integers.
For $\Re r <1$, $I(r)$ is given by Theorem \ref{hsu}, i.e.
$I(r) = H(r)$, where $H$ is the explicit meromorphic function of $r$
(identically 0 if the conditions (\ref{h.4}) are not satisfied)
defined in the preceding subsection.
Therefore, by analytic continuation,
\beq
K(r) \bydef \int_{\CC_1} (1-x^2)^{-r}\,\Phi(x,\ r)\,dx =
(1-e^{-2i\pi r})\,H(r)
= 2i e^{-i\pi r}\sin(\pi r)\,H(r)
\label{c.8}\endq
and
\beq
K'(r) = 2i\pi e^{-2i\pi r} H(r) + (1-e^{-2i\pi r}) H'(r)\ .
\label{c.8.1}\endq
hold for all $r$ at which $H$ is regular.
In particular, for any integer $p$ such that $n_j-2r \ge 0$
for at least two distinct $j$ (see Remark \ref{hrem} above),
\beq
K(p) = 0,\ \ \ \ \ \
K'(p) = 2i\pi H(p)\ .
\label{c.9}\endq
On the other hand
\beq
K'(r) = \int_{\CC_1} \Big [-(1-x^2)^{-r}\,\log(1-x^2)\Phi(x,\ r)
+(1-x^2)^{-r}\,{\partial\over \partial r}\Phi(x,\ r)\Big ]\,dx\ .
\label{c.10}\endq
If $\Phi(x,\ r$ is given by (\ref{c.6}),
\beq
{\partial\over \partial r}\Phi(x,\ r) =
\sum_{\pi} (({\partial/ \partial r})C_{n_j}^{\half-r}(x))\,
C_{n_k}^{\half-r}(x)\, C_{n_l}^{\half-r}(x)\ ,
\label{c.11}\endq
where the sum is over the cyclic permutations $(j,\ k,\ l)$
of $(1,\ 2,\ 3)$.
Let $p$ be  a non-negative integer such that two of the inequalities
$n_1-2p \ge 0$,
$n_2-2p \ge 0$, $n_3-2p \ge 0$, hold. Then (see (\ref{c.11}))
$\Phi(x,\ p)$ and $(\partial/ \partial r)\Phi(x,\ p)$
are polynomials in $x$ divisible by $(1-x^2)^p$.
Therefore the integrands of (\ref{c.8}) and (\ref{c.10}) become,
for $r=p$, integrable on $[-1,\ 1]$, and $K'(r)$ can be
expressed in terms of the integral of its integrand on $[-1,\ 1]$.
The contributions of the two segments of the contour are :
\begin{align}
a\ :\ & \int_{-1}^1 \Big [-(1-x^2)^{-p}\,\log(1-x^2)\Phi(x,\ p)
+(1-x^2)^{-p}\,{\partial\over \partial r}\Phi(x,\ p)\Big ]\,dx\ ,\cr
b\ :\ & -\int_{-1}^1 \Big [ -(1-x^2)^{-p}\,[\log(1-x^2) + 2i\pi]
\Phi(x,\ p) +(1-x^2)^{-p}\,{\partial\over \partial r}\Phi(x,\ p)\Big ]\,dx\ .
\label{c.12}\end{align}
so that
\beq
K'(p) = 2i\pi\,\int_{-1}^1 (1-x^2)^{-p}\,\Phi(x,\ p)\,dx\ ,
\label{c.13}\endq
and hence
\beq
\int_{-1}^1 (1-x^2)^{-p}\,\Phi(x,\ p)\,dx = H(p).
\label{c.14}\endq

\subsection{Extension of Hs\"u's Theorem}

We have thus obtained the following extension of Hs\"u's Theorem:

\begin{lemma}
\label{hsuext}
Let $n_1$, $n_2$, $n_3$ be
non-negative integers and $r \in \bC$
satisfy one of the two following conditions:

\noindent (i) $\Re r < 1$,

\noindent (ii) $r$ is an integer and $n_j -2r \ge 0$ for at least
two distinct values of $j \in \{1,\ 2,\ 3\}$.

Then the statement of Theorem \ref{hsuthm} holds, i.e.
i.e. the integral in the lhs of Eq. (\ref{h.5}) exists and it is equal
to $H(r\ ;\ n_1,\ n_2,\ n_3)$, i.e. the expression (\ref{h.5}) if
the conditions (\ref{h.4}) are satisfied and 0 otherwise.

\end{lemma}

The above phenomenon also occurs if $\Phi(x,\ r)$ is a product
of two Gegenbauer polynomials instead of three. In this case
the role of Hs\"u's formula is played by the orthogonality
relation (see \cite[3.15.1 p. 177]{HTF1}):
\beq
\int_{-1}^1(1-x^2)^{-r}\,C_n^{\half-r}(x)C_m^{\half-r}(x)\,dx
= {2^{2r}\pi\Gamma(n+1-2r)\over (n+\half-r)\Gamma(\half-r)^2
\Gamma(n+1)}\,\delta_{nm}\ ,\ \ \ \Re r < 1\ .
\label{c.15}\endq
The preceding argument, with $\Phi(x,\ r) = C_n^{\half-r}(x)C_m^{\half-r}(x)$,
shows that (\ref{c.15}) still holds for positive integer $r$ such
that $n-2r\ge 0$, $m-2r \ge 0$.

\section{Appendix. An identity for hypergeometric functions}
\label{hypid}
The identity \cite[2.1.5 (26), p.~65]{HTF1}  can be rewritten as
\begin{align}
&F \left (a,\ a+\half\ ;\ c\ ;\ u \right ) =
(v+1)^{2a}F \left (2a,\ 2a-c+1\ ;\ c\ ;\ v \right )\ ,
\label{d.1}\\
&v = {1-(1-u)^\half\over 1+(1-u)^\half}\ ,\ \ \ \ \
u = {4v \over (v+1)^2}\ .
\label{d.2}\end{align}
Letting $z = \half(\zeta+\zeta^{-1})$, $u = z^{-2}$ and $v = \zeta^{-2}$
implies that (\ref{d.2}) holds, and we obtain the identity
\begin{align}
&F \left (a,\ a+\half\ ;\ c\ ;\ {1\over z^2} \right )
= \left (1+{1\over \zeta^2} \right )^{2a}
F \left (2a,\ 2a-c+1\ ;\ c\ ;\
{1\over \zeta^2 } \right )\ ,
\label{d.3}\\
&\zeta = z+(z^2-1)^\half,\ \ \ \zeta^{-1} = z-(z^2-1)^\half,\ \ \ \
z = {\zeta+\zeta^{-1}\over 2}\ .
\label{d.4}\end{align}
Note that (\ref{d.1}) and (\ref{d.3}) can be respectively rewritten as
\beq
u^{a}F \left (a,\ a+\half\ ;\ c\ ;\ u \right )
= (4v)^{a} F \left (2a,\ 2a-c+1\ ;\ c\ ;\ v \right )
\label{d.5}\endq
and
\beq
(2z)^{-2a}F \left (a,\ a+\half\ ;\ c\ ;\ {1\over z^2} \right )
= \zeta^{-2a} F \left (2a,\ 2a-c+1\ ;\ c\ ;\ {1\over \zeta^2 } \right )\ .
\label{d.6}\endq

\section{Appendix. Proof of the projector identity (Theorem \ref{projid})}
\label{adsproj}

In this appendix, $d\ge 2$ is always an integer, and
$\lambda = (d-1)/2$. Hence, for integer $n> -2\lambda$, $D_n^{\lambda}$
is holomorphic in $\wh \Delta_1$.
We will use a very crude bound on $|D_n^\lambda(z)|$ which is valid
if $n$ is real (not necessarily integer or positive),
and $n+2\lambda > 0$, $z \notin [-1,\ 1]$\ :
\beq
|D_n^\lambda(z)| \le {\rm Const.\ } |z|^{-n-2\lambda}
(1+{\rm dist.\,} (z,\ [-1,\ 1])^{-1})^{2\lambda}\ .
\label{w.1}\endq
The constant depends on $n$ and $\lambda$.
This is easily derived from (\ref{s.12}).

It is proved in \cite{brosads} that the tuboid $\TT_{1+}$
(defined in (\ref{s.3})) is the set of points $w=\Lambda z$
where $\Lambda \in G_0$ and $z$ is of the special form
\beq
z = \exp(i\alpha M_{0d}) e_d =
(\sin(i\alpha),\ \vec{0},\ \cos(i\alpha)),\ \ \ \alpha >0\ .
\label{w.2}\endq
If $z \in \TT_{1\pm}$ and $u \in X_d$ then
$z\cdot u \in \bC\setminus [-1,\ 1]$ so that, for integer $n$,
$u \mapsto D_n^\lambda(z\cdot u)$ is $\CC^\infty$ (actually
analytic). We will prove
\begin{lemma}
Let $n_1$, $n_2$ and $d$ be integers satisfying $d \ge 2$,
$n_1+d-1 >0$, $n_2+d-1>0$ and $n_1+n_2+d-1>0$. Let $z_1 \in \TT_{1-}$
and $z_2 \in \TT_{1+}$.
Then the integral
\beq
I_{n_1,n_2,{d-1\over 2}}(z_1,\ z_2) =
\int_{X_d} D_{n_1}^{d-1\over 2}(z_1\cdot u)\,
D_{n_2}^{d-1\over 2}(u\cdot z_2)\,du
\label{w.3}\endq
is absolutely convergent and
\beq
I_{n_1,n_2,{d-1\over 2}}(z_1,\ z_2) =
\delta_{n_1 n_2}\,C(n_1,\ d) D_{n_1}^{d-1\over 2}(z_1\cdot z_2)
\label{w.4}\endq
with
\beq
C(n,\ d) = {4\pi^{d+3\over 2}\over
(2n+d-1) \Gamma \left ({d-1\over 2} \right )}\ .
\label{w.5}\endq
Here $du$ denotes the
standard invariant measure on $X_d$, i.e.
$du = 2\delta(u\cdot u -1)du^0\ldots du^d$. The convergence takes place
in the space of functions with tempered behavior in
$\TT_{1-}\times \TT_{1+}$.
\end{lemma}
Note that if $n_1=n_2=n$, the condition $2n+d-1>0$ must be satisfied.

\noindent {\bf Proof.} Let $n$ and $d$ be integers such that
$d \ge 2$ and $n+d-1 >0$ and $p \in \bR$ satisfy $p(n+ d-1)-d+1>0$,
i.e.
\beq
0 < {1\over p} < {n+d-1 \over d-1}\ .
\label{w.9}\endq
We will verify that, for $z=x+iy \in \TT_+$ (and similarly for $z \in \TT_-$),
the integral
\beq
J_{n,p}(z) = \int_{X_d} |D_n^{d-1\over 2}(z\cdot u)|^p\,du
\label{w.6}\endq
is absolutely convergent. By a transformation in $G_0$ we can
bring $z$ to the form (\ref{w.2}).
Taking $u$ in the form
\beq
u = (s\sin\theta,\ r\vec{v},\ s \cos\theta)
\label{w.7}\endq
with $\vec{v}$ a $(d-1)$-dimensional unit vector, we find after
integrating over $\vec{v}$
\begin{align}
J_{n,p}(z) &= 2\Omega_{d-1}\int_0^\infty s\,ds\int_0^\infty r^{d-2}\,dr
\int_0^{2\pi}d\theta\,\delta(s^2-r^2-1)\,
|D_n^{d-1\over 2}(s\cos(\theta-i\alpha))|^p\cr
&= \Omega_{d-1}\int_0^\infty r^{d-2}\,dr\int_0^{2\pi}
|D_n^{d-1\over 2}(\sqrt{r^2+1}\cos(\theta-i\alpha))|^p\,d\theta\ ,\ \ \ \
\Omega_{d-1} = {2\pi^{d-1\over 2}\over \Gamma \left ({d-1\over 2} \right )}.
\label{w.8}\end{align}
At fixed $r$, the argument $\tau$ of $D_n^{d-1\over 2}$ follows the ellipse
$\sqrt{r^2+1}\,E(e^\alpha)$ with foci $\pm \sqrt{r^2+1}$ and minor
semi-axis $\sqrt{r^2+1}\sh(\alpha)$. Hence
$|\tau| \ge \sqrt{r^2+1}\sh(\alpha)$ and
${\rm dist.\,}(\tau,\ [-1,\ 1]) \ge \sqrt{r^2+1}(\ch(\alpha)-1)$.
The full integrand is thus majorized by
${\rm const.\ }(\sh \alpha)^{-p(n+d-1)}r^{-p(n+d-1)+d-2}
(1+\sh(\alpha/2)^{-2})^{pd}$,
and the integral is absolutely convergent. Since $y\cdot y = \sh(\alpha)^2$,
the convergence takes place in the space of functions of $z$ bounded
in modulus by a fixed negative power of  $y\cdot y$.

The condition $n_1+n_2 +d-1 >0$ postulated in the lemma can
be rewritten as
\beq
1< {n_1+d-1 \over d-1} + {n_2+d-1 \over d-1}\ .
\label{w.10}\endq
If it is satisfied, it is possible to find $p_1 >0$, $p_2>0$ such that
\beq
{1\over p_1}+ {1\over p_2} = 1,\ \ \
{1\over p_1}< {n_1+d-1 \over d-1},\ \ \
{1\over p_2}< {n_2+d-1 \over d-1}\ .
\label{w.11}\endq
By H\"older's inequality,
\beq
|I_{n_1,n_2,{d-1\over 2}}(z_1,\ z_2)| \le
(J_{n_1,p_1}(z_1))^{1\over p_1}\,(J_{n_2,p_2}(z_2))^{1\over p_2}\ ,
\label{w.13}\endq
in which each of the integrals in the rhs is absolutely convergent.
Again the convergence takes place in the space of functions with
tempered behavior in $\TT_{1-}\times \TT_{1+}$. Since the convergence
is in particular uniform on compact subsets of $\TT_{1-}\times \TT_{1+}$,
the result
$I_{n_1,n_2,{d-1\over 2}}(z_1,\ z_2)$ is an invariant function of $z_1$ and $z_2$,
holomorphic in $\TT_{1-}\times \TT_{1+}$. It is therefore
equal to a function of
$z_1\cdot z_2$ holomorphic in $\Delta_1$ and tending to 0 at infinity,
thus holomorphic in $\wh \Delta_1$. It satisfies,
in each of the two variables, the Klein-Gordon equation
with square masses $n_1(n_1+d-1)$ and $n_2(n_2+d-1)$ respectively.
It must therefore vanish if $n_1 \not= n_2$. If $n_1=n_2$,
these properties characterize
$w_{n_1+{d-1\over 2}}(z_1\cdot z_2)$ up to a constant factor.
It follows that (\ref{w.4}) holds.
In order to compute the constant $C(n,\ d)$, we specialize
to the case $n_2 = n_1 = n$ and $z_1^* = z_2 =z$, $z$ of the form
(\ref{w.2}). We find
\beq
\Omega_{d-1}\int_0^\infty r^{d-2}\,dr\int_0^{2\pi}
|D_n^{d-1\over 2}(\sqrt{r^2+1}\cos(\theta-i\alpha))|^2\,d\theta =
C(n,\ d) D_{n}^{d-1\over 2}(\ch(2\alpha))\ .
\label{w.14}\endq
Setting $\zeta = e^\alpha$, both sides are analytic in $\zeta$ in
a neighborhood of infinity and have a convergent expression
in $1/\zeta$. Identifiying the first terms of the expansions
of both sides will give $C(n,\ d)$:
\beq
\zeta^{-2n-2d+2}{\Omega_{d-1}2\pi^2\Gamma(n+d-1)\over
\Gamma \left ({d-1\over 2} \right )\Gamma(n+{d+1\over 2})} \int_0^\infty
r^{d-2}(r^2+1)^{-n-d+1}\,dr =
C(n,\ d)\zeta^{-2n-2d+2}\ ,
\label{w.15}\endq
\beq
C(n,\ d) = {\Omega_{d-1}2\pi^2\Gamma(n+d-1)\over
\Gamma \left ({d-1\over 2} \right )\Gamma \left ( n+{d+1\over 2} \right )}
\int_0^\infty r^{d-2}(r^2+1)^{-n-d+1}\,dr \ .
\label{w.16}\endq
The last integral is equal to
\begin{align}
{1\over 2}\int_0^\infty t^{d-3\over 2}(t+1)^{-n-d+1}\,dt
&= {1\over 2} \Beta \left ({d-1\over 2},\ n+{d-1\over 2} \right )\cr
&= {\Gamma \left ({d-1\over 2} \right )
\Gamma \left (n+{d-1\over 2} \right ) \over
2\Gamma(n+d-1)}
\label{w.17}\end{align}
(\cite[p.~9]{HTF1}). Hence
\beq
C(n,\ d) = {4\pi^{d+3\over 2}\over
(2n+d-1) \Gamma \left ({d-1\over 2} \right )}\ .
\label{w.18}\endq
This completes the proof of the lemma.

We may reexpress this in terms of $w_\nu$, $\nu = n+(d-1)/2$, given
by (\ref{s.10}):
\beq
\int_{X_d} w_{n_1+{d-1\over 2}}(z_1\cdot u)\,
w_{n_2+{d-1\over 2}}(u\cdot z_2)\,du =
{2\pi \over (2n_1+d-1)}\delta_{n_1n_2} w_{n_1+{d-1\over 2}}(z_1\cdot z_2)\ .
\label{w.19}\endq
This proves Theorem \ref{projid}.

\end{document}